\documentclass[]{aa}
\pdfoutput=1                            
\usepackage[varg]{txfonts}

\usepackage{amsmath}
\usepackage{amsfonts}
\usepackage{amssymb}
\usepackage{wasysym}
\usepackage{graphicx}

\usepackage{natbib}
\usepackage{url}
\usepackage{breakurl}
\usepackage[breaklinks=true]{hyperref}
\usepackage{epsfig}
\usepackage{xspace}
\usepackage{booktabs}
\usepackage{dcolumn}
\usepackage[para]{footmisc}
\usepackage{placeins}
\usepackage{multicol}
\usepackage{color}
\usepackage[switch]{lineno}

\usepackage[utf8]{inputenc}
\usepackage[english]{babel}

\newcommand{\teff}{\ensuremath{T_\mathrm{Eff}}\xspace}
\newcommand{\teffh}{\ensuremath{T_\mathrm{Eff,H}}\xspace}
\newcommand{\teffc}{\ensuremath{T_\mathrm{Eff,C}}\xspace}
\newcommand{\pzep}{\ensuremath{t_\mathrm{0}}\xspace}
\newcommand{\pper}{\ensuremath{p}\xspace}
\newcommand{\prho}{\ensuremath{\rho_\star}\xspace}
\newcommand{\pimp}{\ensuremath{b}\xspace}

\newcommand{\pcref}{\ensuremath{c_\mathrm{i'}}\xspace}
\newcommand{\para}{\ensuremath{k_\mathrm{app}}\xspace}
\newcommand{\ptra}{\ensuremath{k_\mathrm{true}}\xspace}
\newcommand{\paaa}{\ensuremath{k^2_\mathrm{app}}\xspace}
\newcommand{\ptaa}{\ensuremath{k^2_\mathrm{true}}\xspace}

\newcommand{\mjup}{\ensuremath{M_\mathrm{Jup}}\xspace}
\newcommand{\rjup}{\ensuremath{R_\mathrm{Jup}}\xspace}

\newcommand{\pytransit}{\textsc{PyTransit}\xspace}
\newcommand{\ldtk}{\textsc{LDTk}\xspace}
\newcommand{\george}{\textsc{George}\xspace}
\newcommand{\ud}{\ensuremath{\mathrm{d}\xspace}}

\newcommand{\plato}{\textit{PLATO}\xspace}
\newcommand{\tess}{\textit{TESS}\xspace}
\newcommand{\corot}{\textit{CoRoT}\xspace}
\newcommand{\kepler}{\textit{Kepler}\xspace}
\newcommand{\ktwo}{\textit{K2}\xspace}

\begin{document}
   \title{Multicolour photometry for exoplanet candidate validation}
  
   \author{H.~Parviainen\inst{\ref{iiac},\ref{iull}} 
   		\and B.~Tingley\inst{\ref{ibran}}  
   		\and H.~J. Deeg\inst{\ref{iiac},\ref{iull}} 
   		\and E.~Palle\inst{\ref{iiac},\ref{iull}} 
   		\and R.~Alonso\inst{\ref{iiac},\ref{iull}} 
   		\and P.~Montanes Rodriguez\inst{\ref{iiac},\ref{iull}} 
   		\and F. Murgas\inst{\ref{iiac},\ref{iull}} 
   		\and N.~Narita\inst{\ref{iiac},\ref{iutda},\ref{iabc},\ref{ijsta},\ref{inao}} 
   		\and A.~Fukui\inst{\ref{idepsut},\ref{iiac}}
   		\and N.~Kusakabe \inst{\ref{iabc},\ref{inao}} 
   		\and M.~Tamura\inst{\ref{iutda},\ref{iabc},\ref{inao}}
   		\and T.~Nishiumi\inst{\ref{ikyo}}
   		\and J.~Prieto-Arranz\inst{\ref{iiac},\ref{iull}}
   		\and P.~Klagyivik\inst{\ref{iiac},\ref{iull}}
   		\and V.~J.~S.~B\'ejar\inst{\ref{iiac},\ref{iull}}
   		\and N.~Crouzet\inst{\ref{iesa}}
   		\and M.~Mori\inst{\ref{iutda}}
   		\and D.~Hidalgo~Soto\inst{\ref{iiac},\ref{iull}}
   		\and N.~Casasayas~Barris\inst{\ref{iiac},\ref{iull}}
   		\and R.~Luque\inst{\ref{iiac},\ref{iull}}
   		}

    \institute{
  	     Instituto de Astrof\'isica de Canarias (IAC), E-38200 La Laguna, Tenerife, Spain\label{iiac}
  	\and Dept. Astrof\'isica, Universidad de La Laguna (ULL), E-38206 La Laguna, Tenerife, Spain\label{iull}
  	\and B{\o}ggildsvej 14, 8530 Hjortsh{\o}j, Denmark\label{ibran}
  	\and Department of Astronomy, The University of Tokyo, 7-3-1 Hongo, Bunkyo-ku, Tokyo 113-0033, Japan \label{iutda}
  	\and Astrobiology Center, 2-21-1 Osawa, Mitaka, Tokyo 181-8588, Japan \label{iabc}
  	\and Japan Science and Technology Agency, PRESTO, 2-21-1 Osawa, Mitaka, Tokyo 181-8588, Japan \label{ijsta} 
  	\and National Astronomical Observatory of Japan, 2-21-1 Osawa, Mitaka, Tokyo 181-8588, Japan \label{inao}
  	\and Department of Earth and Planetary Science, Graduate School of Science, The University of Tokyo, 7-3-1 Hongo, Bunkyo-ku, Tokyo 113-0033, Japan \label{idepsut} 
  	\and Subaru Telescope Okayama Branch Office, National Astronomical Observatory of Japan, Okayama, Japan \label{isub}
  	\and SOKENDAI (The Graduate University of Advanced Studies), Tokyo, Japan \label{isok}
  	\and  Department of Physics, Kyoto Sangyo University, Motoyama, Kamigamo, Kita-ku, Kyoto, 603-8555 Japan \label{ikyo}
    \and European Space Agency, European Space Research and Technology Centre, Keplerlaan 1, 2201 AZ Noordwijk, The Netherlands \label{iesa}
  }
    \date{Received December XX, 2017; accepted YYY XX, 2017}
 
  \abstract 
  {The \tess and \plato missions are expected to find vast numbers of new transiting planet candidates. 
    However, only a fraction of these candidates will be legitimate planets, and the
  	candidate validation will require a significant amount of follow-up resources.
  	Radial velocity (RV) follow-up can be carried out only for the most promising candidates around bright,
  	slowly rotating, stars. Thus, before devoting RV resources to candidates, they need to be vetted
  	using cheaper methods, and, in the cases for which an RV confirmation is not feasible, the
  	candidate's true nature needs to be determined based on these alternative methods alone.}
  {We study the applicability of multicolour transit photometry in the validation of transiting 
    planet candidates when the candidate signal arises from a real astrophysical source (transiting planet,
    eclipsing binary, etc.), and not from an instrumental artefact. Particularly, we aim to answer how 
    securely can we estimate the true uncontaminated star-planet radius ratio when the light
    curve may contain contamination from
    unresolved light sources inside the photometry aperture when combining multicolour transit
    observations with a physics-based contamination model in a Bayesian parameter estimation setting. More 
    generally, we study how the contamination level, colour differences between the planet host and 
    contaminant stars, transit signal-to-noise ratio, and available prior information 
  	affect the contamination and true radius ratio estimates.} 
 {The study is based on simulations and ground-based multicolour transit observations. The contamination
 	analyses are carried out with a contamination model integrated into the \pytransit~v2 transit modelling
 	package, and the observations are carried out with the MuSCAT2
 	multicolour imager installed in the 1.5~m Telescopio Carlos Sanchez in the Teide Observatory, Tenerife.}
 {We show that multicolour transit photometry can be used to estimate the amount of flux contamination
    and the true radius ratio. Combining the true radius ratio with an estimate for the stellar radius yields 
    the true absolute radius of the
 	transiting object, which is a valuable quantity in statistical candidate validation, and enough in itself
 	to validate a candidate whose radius falls below the theoretical lower limit for a brown dwarf.}{}
  \keywords{ Planetary systems -- planets and satellites: detection -- methods: statistical -- methods: numerical -- methods: data analysis -- techniques: photometric}

  \maketitle

\section{Introduction}

Both the currently operational \tess and the upcoming \plato
mission are expected to find vast numbers of new exoplanet candidates
\citep{Ricker2014,Rauer2013}, while the current ground-based surveys, such as
\textit{WASP} \citep{Pollacco2006}, continue producing new
discoveries on a steady pace. However, as with the previous \corot, \kepler, and \ktwo missions,
only a fraction of these candidates will be legitimate planets
\citep{Moutou2009,Almenara2009,Santerne2012,Fressin2013}. Several astrophysical
phenomena--such as eclipsing binaries (EBs), blended eclipsing binaries (BEBs), and
transiting brown dwarfs--can lead to a photometric signal that mimics an exoplanet transit
\citep{Brown2003,Cameron2012}, and most of the candidates require follow-up observations to clarify
the nature of the object causing the observed transit-like signal \citep[e.g.,][]{Cabrera2017a,Mullally2018}.

Measuring the mass of a transiting object using radial velocity (RV) observations is
considered to be the most reliable technique for the confirmation of transiting exoplanet candidates.
However, RV observations can be carried out only with a small number of specialised
instruments installed in high-demand telescopes,  and the observations are restricted to
candidates with a relatively high planet-star mass ratio orbiting  bright host stars.
Moreover, RVs can be difficult to measure for certain types of stars--for example fast
rotators, hot stars, or very metal poor stars--due the absence or broadening of the
absorption lines on which the cross-correlation used to measure the RV depends.

While the methods available for planet candidate vetting and validation\footnote{Here we
consider candidate vetting as a procedure that can identify certain types of false
positives, while candidate validation aims to statistically validate the candidate as a
planet using information from (preferably) multiple vetting methods.}  have diversified
during the last decade,$\!$\footnote{Current standard methods include the use of the original
space-based photometry
\citep{Batalha2010,Quintana2013,Bryson2013,Coughlin2014,Mullally2016,Armstrong2017},
ground-based photometry and high-resolution imaging \citep{Deeg2009,Guenther2013}, and
statistical candidate validation using evidence from multiple information sources, such as
\textsc{Blender} \citep{Torres2011}, \textsc{PASTIS} \citep{Diaz2014,Santerne2015}, and
\textsc{vespa} \citep{Morton2016}.} there is still a strong demand for observationally
economical approaches to reduce the amount of false positives reaching the RV follow-up
phase. Additionally, methods that can reliably identify false positives are required in
candidate validation for cases where the RV follow-up is not viable, such as with rapidly
rotating host stars.

Eclipsing binaries, and in particular blended eclipsing binaries, where the light of an
unresolved EB contaminates an otherwise constant star \citep{Brown2003,Mandushev2005,Cameron2012}, can
closely resemble exoplanet transits. BEBs are a common source of false positive transit
signals, and can be problematic for RV analysis. For example, a faint EB
blended with a bright star may not show any detectable RV signal since the bright star
dominates the spectrum. Another (although rarer) case is that of a BEB where all the stars 
have approximately the same colour. This happened with WASP-9b,
where the system exhibited strong radial velocity variations consistent with a planet with
a mass of $2.3\,\mjup$, showed no sign of any bisector variation,  yet still proved
to be a false positive.$\!$\footnote{ When the WASP team attempted to detect the
	Rossiter-McLaughlin effect for this already-announced exoplanet, they noticed highly
	rotationally-broadened spectral lines shifting back and forth with the same period as
	their proposed exoplanet. The only possible explanation for this phenomena was that
	WASP-9b was actually not a planet, but a BEB with little colour difference between the
	stars. Generally, the presence of a transit combined with a mass from RV measurements that
	exhibit no bisector variations is considered enough for confirmation, yet the case of
	WASP-9b shows that these criterion may not be sufficient in all cases.}

Both \tess and \plato have large pixel sizes (21\arcsec and 15\arcsec{}, respectively), which leads to
blending  being a more significant issue than with either \corot or \kepler with pixel sizes of
2.32\arcsec and 3.98\arcsec, respectively. Thus, even
while PSF (point spread function) centroid variations can be used to identify many of the blending cases \citep{Bryson2013},
the \tess and \plato candidate vetting will rely heavily on ground-based photometric
follow up.

This paper explores the use of multicolour transit photometry in transiting exoplanet
candidate vetting and validation, continuing the ideas presented originally by
\citet{Rosenblatt1971}, developed further by \citet{Drake2003} and \citet{Tingley2004},
and applied in practice by \citet{Tingley2014a}.

While the mass of a planet obtained from the RV observations is an important quantity,
it is not strictly necessary for the validation of a planet candidate. Restricting the
radius of the transiting body to planetary size is an equally viable option. This may
proceed by establishing a radius less than the smallest brown dwarfs ($\sim 0.8\, \rjup$,
\citealt{Burrows2011}) or a radius consistent with a hot Jupiter and an upper mass limit (via
RV) less than a brown dwarf ($\sim 13\,\mjup$, \citealt{Chabrier2000}). The radius of a
transiting body can be measured in two different ways, either using the
Rossiter-McLaughlin effect \citep{Worek2000,Gimenez2006} or through the analysis of
transit photometry. The latter yields an estimate for the (apparent) planet-star radius ratio based
on the transit depth, $R_p/R_\star \approx (\Delta F/F_0)^{1/2}$. However, this ratio is
valid only if we can assure that the occulting body is not performing a grazing eclipse or
that the light curve does not contain any relevant flux from a third body (flux contamination,
from now on). If the light curve is contaminated, the measured \emph{apparent} radius ratio
is smaller than the \emph{true} radius ratio (that is, the transit appears shallower than it truly is),
and a radius ratio estimate that has been derived with a model that does not include contamination
can not be trusted.

 Multicolour photometry has already been
used in the vetting of space-based transit candidates, both from \kepler; e.g.  in
combination with data from Spitzer \citep{Ballard2011a} or from precise ground-based photometry
\citep{Colon2011,Tingley2014a}, as well as in data from \corot, where colour differences among its
three-channel data were routinely used to reject planet candidates \citep{Carone2011}.
However, multicolour transit photometry can be utilised further than what is currently
done. For candidate vetting purposes, it allows one to estimate the true radius ratio
that accounts for possible flux contamination from unresolved sources, including the transiting object itself, using 
relatively small (possibly automatised) ground-based telescopes, thus revealing blends.
The true radius ratio estimate can then be combined with the stellar radius estimate to produce
an estimate for the absolute planet radius, which can be used in the candidate validation.
As an additional benefit, the observations improve the candidate's ephemeris,
which can be crucial when following up candidates from missions consisting of relatively
short stares, such as \tess \citep{Deeg2016}. 

In this paper, we study the applicability of multicolour transit photometry in transiting planet 
candidate validation, assuming that the candidate signal arises from a real astrophysical event
(such as a planetary transit or binary eclipse), and is not from an instrumental source.
Particularly, we aim to answer how accurately can the true radius ratio 
be estimated, and how the contamination level, host-contaminant 
colour difference, transit signal-to-noise ratio, and available prior information 
affect the contamination and true radius ratio estimates.
  	
Most previous work involving multicolour photometry of planetary transits has
interpreted the different passbands individually; e.g. by fitting a transit model on each
passband individually. In this work, we combine a physics-based contamination model with
a Bayesian parameter estimation approach where 
all passbands are modelled jointly. This allows us to break the degeneracies between the
impact parameter, radius ratio, stellar limb darkening, and possible flux
contamination, yielding improved orbital parameter estimates, an estimate for the level of
contamination, and a robust true planet-star radius ratio estimate.

The simulation and light curve analysis codes developed for this paper are publicly
available from GitHub. The contamination model is detailed in
Appendix~\ref{sec:appendix.model}, and included in the transit modelling package
\pytransit v2, which is also publicly available from GitHub\footnote{\url{https://www.github.com/hpparvi/pytransit}}.

\section{The transit colour signatures}
\label{sec:colour_signature}

Flux contamination (blending) decreases the observed transit depth and leads to incorrect
transit--and thus planetary--parameters \citep{Daemgen2009}. The effects from
contamination on a single passband are degenerate with the effects from orbital geometry,
stellar limb darkening, and radius ratio, which means single colour observations cannot 
generally be used to constrain contamination. 

Two separate colour-dependent effects yield information about the degree of contamination and the 
true radius ratio of the transiting object:
\begin{enumerate} 
	\item Colour differences between the stars contributing flux to the
	observed light curve will lead to variations in the transit depth observed in different
	passbands \citep{Drake2003,Tingley2004}. This makes identifying EBs and BEBs relatively
	easy, since the variations in transit depth can be significant if the stars have very
	different colours. 
	\item  The transit itself produces a colour-dependent signal where the shape of the signal
	depends on the size of the transiting object. As originally noted by \citet{Rosenblatt1971} and elaborated in
	more detail by \citet{Tingley2004}, systems with ratio of radii genuinely consistent with
	exoplanets exhibit a distinctive, double-horned colorimetric signature during transit that
	increases in prominence relative to the transit depth as the radius ratio decreases.
	This effect does not depend on the colour differences between the host and the contaminant(s), and can be 
	used to estimate blending even when all the components have the same colour. 
\end{enumerate}
 Contamination measures based on the analysis of multicolour photometric time series can
reveal all contaminating sources, regardless of proximity in the sky -- including those
that are actually gravitationally bound to the host star, and therefore extremely hard to resolve.

\begin{figure}
	\centering 
	\includegraphics[width=\columnwidth]{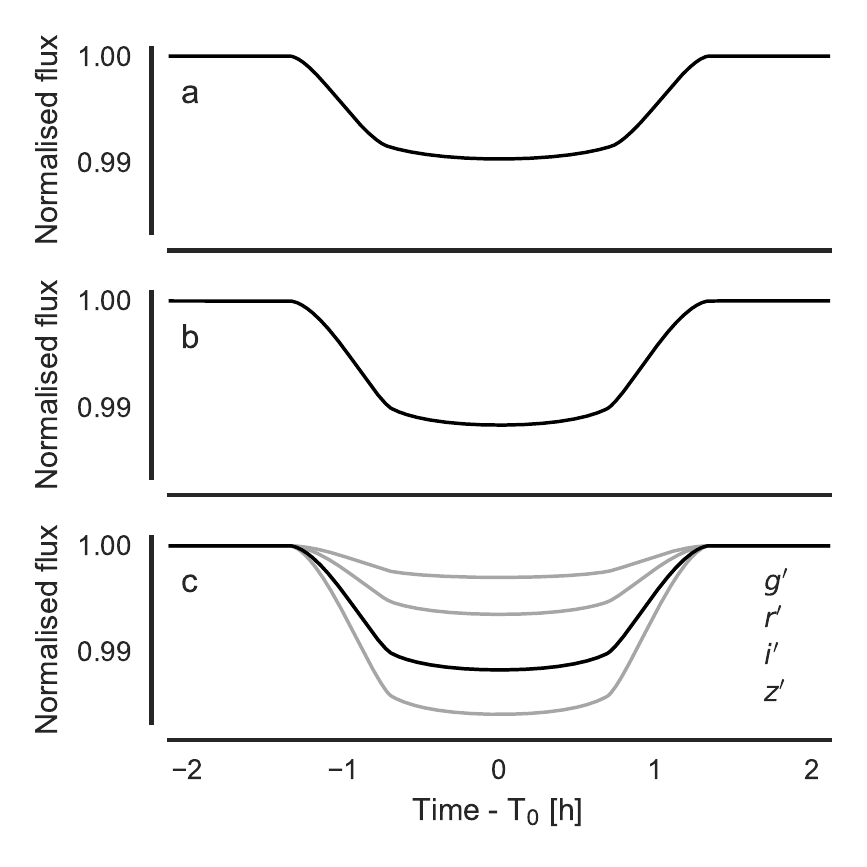}
	\caption{An uncontaminated transit by a planet with radius ratio of 0.1 observed in the $i'$ band (a), a faint 
		M-dwarf transited by an object with a radius ratio of 0.32 strongly contaminated by a G star ($c=0.9$, i.e., 10\% of the 
		total flux comes from the M dwarf, and 90\% from the G star, )
		leading to an apparent radius ratio of 0.1 in the $i'$ band (b), and the same contaminated transit observed in 
		$g'$, $r'$, $i'$, and $z'$ bands (c).} 
	\label{fig:lc_comparison_1}
\end{figure}

\begin{figure*}
	\centering 
	\includegraphics[width=\textwidth]{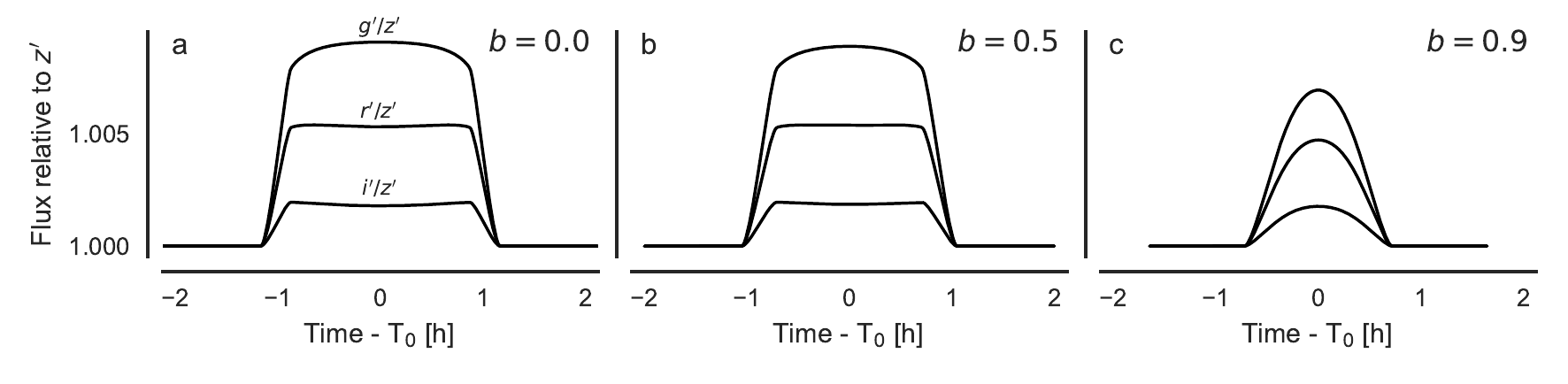}
	\caption{Effect of varying impact parameter on the transit depth colour differences when the host and contaminant stars are of different spectral types (M and G). The setup is the same as in Fig.~\ref{fig:lc_comparison_1} panel~c, but the 
	impact parameter varies from panel to panel.
	The colour differences between the stars lead to differences in the relative contamination from one passband to another,
	which leads to variations in the observed transit depths.} 
	\label{fig:lc_comparison_2}
\end{figure*}

The first effect, colour-dependent transit depth variations due to contamination from
a star of a different spectral type (colour) than the host star, is well known and has been used in planet
candidate validation \citep{ODonovan2006,ODonovan2007,Ballard2011a,Cochran2011a,Tingley2011b,Colon2011}.
Figure \ref{fig:lc_comparison_1} demonstrates the colour-dependent transit depth
variations when the host and contaminant stars have different colours. Panel a. shows an
uncontaminated transit by a planet with a \emph{true planet-star radius ratio} (\ptra) of 0.1 observed in the
$i'$ band, and panel b. an M-dwarf eclipsed by an object with a true radius ratio of 0.32 blended with
a G-star so that 90\% of the total flux comes from the contaminant ($c=0.9$) leading to
an \emph{apparent radius ratio} (\para) of 0.1 in the $i'$ band. Panel c. shows the same contaminated transit in
$g'$, $r'$, $i'$, and $z'$, where the strong transit depth variations make the significant
contamination evident. Figure \ref{fig:lc_comparison_2}  demonstrates the transit depth
variations further as a function of impact parameter. Since the variations are due to the
changing contamination (i.e., the relative brightness of the stars vary as a function of
wavelength), the effect is not dependent on the geometric properties of the transiting
object's orbit.

\begin{figure*}
	\centering 
	\includegraphics[width=\textwidth]{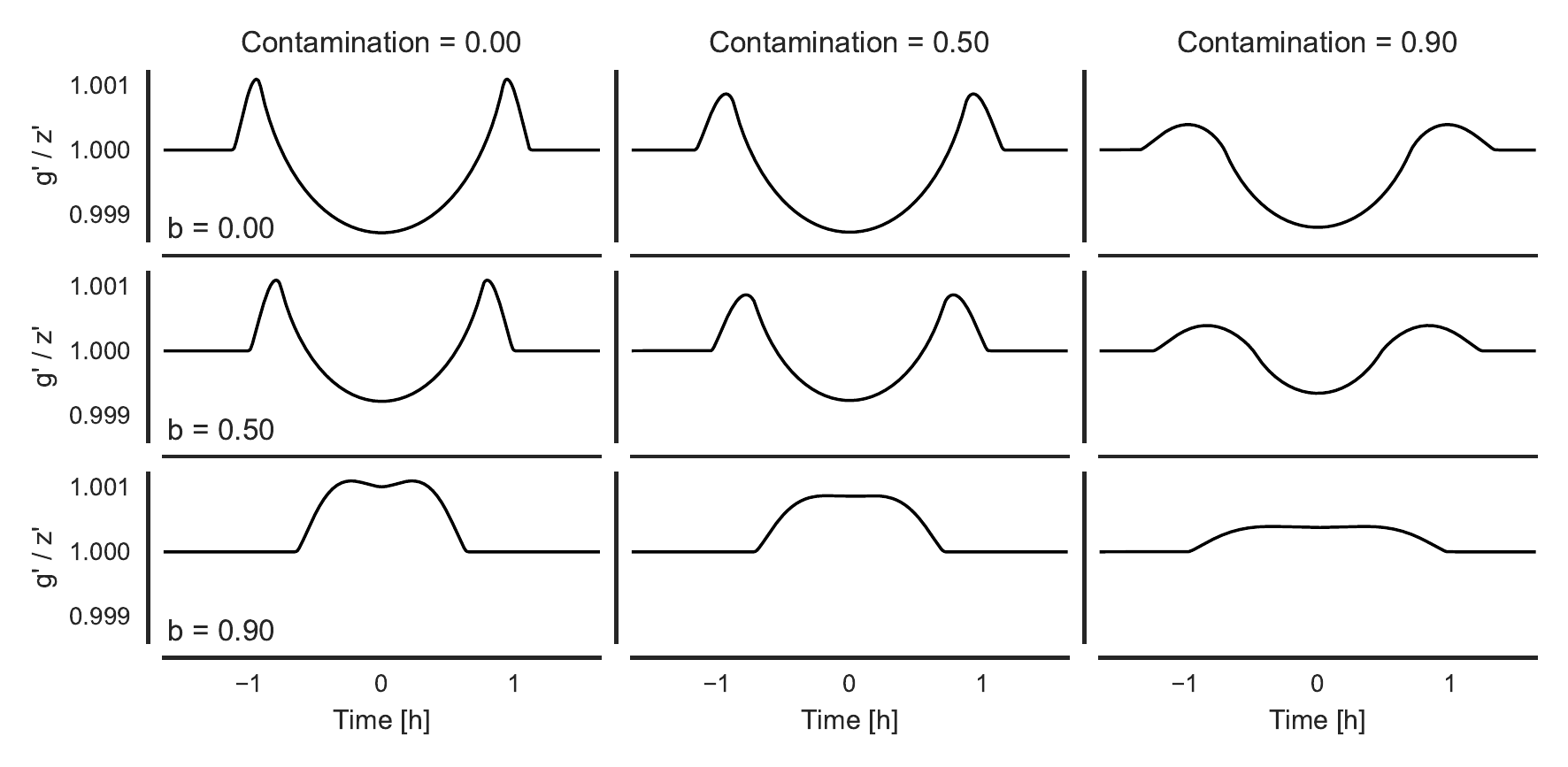}
	\caption{Examples of the transit colour differences when the host and contaminant are of same spectral type for three
	contamination values and three impact parameters. The smaller the transiting object is, the sharper the 
	differential features are.} 
	\label{fig:lc_comparison_3}
\end{figure*}

The second effect is less well known, and is illustrated in Fig.~\ref{fig:lc_comparison_3}.
\citet{Rosenblatt1971} was the first to propose to use the distinctive signature that
appears in multicolour time series photometry of the parent star during an exoplanet
transit to discriminate between eclipsing binaries and transiting exoplanets. This
signature arises from the interplay between the relatively small size of a planet compared
to its parent star and differential limb darkening. Qualitatively speaking, the light
coming from a stellar disk is bluer in its centre than near its limb. Therefore, at the
beginning of all uncontaminated eclipses and transits, the integrated light from the
stellar disk becomes bluer, as the redder light at the limb is occulted. What follows
depends primarily on the radius-ratio of the system \citep[see][for details]{Tingley2004}.
At one extreme, if the transiting object is much smaller than the transited one, the redder
light at the limb is revealed and the bluer light at the centre of the disk is occulted as
the transit proceeds, causing a red-ward shift in the integrated light. This reaches a
maximum at transit centre, after which the entire process is repeated in reverse during
egress. This results in a sharp blue spike in the colorimetry at both ingress and egress,
with a red-ward bulge near the transit centre, the details of which depend on the impact
parameter. The situation at the other extreme -- two bodies approximately the same size --
is markedly different, as the redder light at the limb remains occulted for most or all of
the transit, resulting in a distinct absence of sharp features of any sort -- whether or
not light from any third star is included.

In practice, these two effects can be included in the transit model used to
model the multicolour photometry. Including 
a passband dependent contamination into a transit model takes both effects into account
naturally, and allows for the per-passband contamination and the true radius ratio to be 
estimated directly from the multicolour observations.

\section{Numerical methods}
\label{sec:methods}

We developed a physics-based contamination model that is used to carry out the
analyses presented in this paper, and included it in  \pytransit~v2
\citep{Parviainen2015}. The model uses synthetic stellar spectra calculated with
 \textsc{PHOENIX} by \citet{Husser2013}, and is detailed in Appendix~\ref{sec:appendix.model}.

The analysis of simulations and observations is carried out with \textsc{Python} code
utilising \textsc{SciPy}, \textsc{NumPy} \citep{VanderWalt2011},  \textsc{AstroPy}
\citep{TheAstropyCollaboration2013}, \textsc{IPython} \citep{Perez2007}, \textsc{Pandas}
\citep{Mckinney2010}, \textsc{matplotlib} \citep{Hunter2007},
\textsc{seaborn},$\!$\footnote{\url{http://stanford.edu/~mwaskom/software/seaborn}} and
\textsc{F2PY} \citep{Peterson2009}. The limb darkening computations were carried out with
\ldtk$\!$\footnote{\url{https://github.com/hpparvi/ldtk}} \citep{Parviainen2015b}, global
optimisation was carried out with
\textsc{PyDE},$\!$\footnote{\url{https://github.com/hpparvi/PyDE}} the MCMC sampling was
carried out with \textsc{emcee} \citep{Foreman-Mackey2012,Goodman2010}, and the Gaussian
processes (GPs) were computed using \george\footnote{\url{https://dan.iel.fm/george}}
\citep{Ambikasaran2014}.

\section{Simulations}
\label{sec:simulations}

\subsection{Overview}
\label{sec:simulations.overview}

Before studying real observations, we study the practical applicability of the multicolour
validation method using simulated transit light curves. The simulations
mimic observations by an instrument observing four passbands simultaneously ($g'$, $r'$,
$i'$, $z'$), installed in a ($\sim 1.5$~m) ground-based telescope located in an
observatory with good observing conditions. The simulations were designed to match closely
the instrument and telescope setup used for the observational study presented in
Sec.~\ref{sec:reality}, that is, MuSCAT2 multicolour imager installed in the 1.5~m Telescopio Carlos Sanchez in the
Teide Observatory \citep{Narita2018}. The noise level, 1000~ppm over an exposure of 60~s, is also based on existing observations for a $V=12$ star with this setup. However, we consider
only white noise, and leave the effects from correlated noise for a later study.

The simulations are divided into separate sets of illustrative scenarios. For each scenario, the computations
are  repeated for combinations of apparent radius ratio ($\para = \{0.07,\, 0.10,\, 0.15\}$), impact parameter
($\pimp = \{0.0,\, 0.5,\, 0.85\}$), number of observed transits ($n_\mathrm{n} = \{1,\, 2,\, 4\}$), and transit 
duration ($T_{14}\sim\{1~\mathrm{h},\, 2~\mathrm{h}\}$), leading to 54 simulations per scenario. 

We parametrise the planet candidate orbit by the zero epoch~(\pzep), period (\pper),
stellar density (\prho), and impact parameter (\pimp), and assume zero eccentricity for simplicity. The
planet and the contamination are parametrised by  the effective
temperatures of the host (\teffh) and contaminant (\teffc), the true uncontaminated
planet-star area ratio (\ptaa), and the apparent area ratio in the $i'$ passband (\paaa). The
contamination in the reference  $i'$ band is calculated from the true and contaminated
area ratios, and \pytransit's contamination module is used to calculate the contamination
in the rest of the passbands.

We carry out most of the simulations both with and without an informative prior on \teffh,
where the informative prior is set to constrain \teffh close to the true value of the
dominant component \teff (that is, we also study cases where the contaminant dominates and
we misidentify the host star).  We set an uninformative prior on \pzep, and a tight normal
prior on \pper (which usually can be assumed to be known well). The two orbital
parameters, \prho and \pimp, have either uninformative priors or tight informative priors
to study the effects of having prior knowledge on the planet candidate orbit. These two
options represent extreme cases. Many of the orbital parameters are degenerate in low
signal-to-noise scenarios, and the uninformative-priors-cases are used to study how the
contamination posterior behaves in a poorly constrained situation. In practise, especially
when following up candidates found by space-based transit surveys, we can obtain
informative priors for \prho and \pimp based on the existing photometry (as long as we
remember to use a transit model that includes contamination also in the analysis of the
existing photometry), or model the existing photometry jointly with the multicolour
photometry. Informative priors based on high signal-to-noise photometry can help to reduce
the degeneracies and improve the reliability of the contamination estimate.

Finally, limb darkening is also degenerate with the impact parameter, radius ratio, and
contamination, and constraining limb darkening could help to reduce the degeneracies.
However, since in a real situation we do not know if the host is the dominant source or
not, we choose a conservative approach and marginalise (average) over the whole limb darkening
coefficient space.

\subsection{M-dwarf without contamination}
\label{sec:simulations.m_no_contamination}

We begin by considering an uncontaminated ($\teff=3600$~K, $\prho=
5\;\mathrm{g\,cm^{-3}}$) M dwarf planet host with and without an informative prior set on the host
star effective temperature (\teffh), and with and without informative priors on the stellar density and impact
parameter. An estimate for \teffh derived from spectroscopy (or from the same multicolour
photometry used in the contamination analysis) helps to constrain the parameter space, as
do estimates for the orbital parameters derived from existing transit photometry.
However, if the contaminant is brighter than the host, the \teffh estimate can be
erroneous. Also, the orbital parameter estimates used as priors need to be derived using a
model that allows for contamination (or the prior photometry needs to be modelled jointly
with the multicolour observations). If the orbital parameter priors are based on an analysis
with a model that does not include contamination, they will most likely be biased and
unrealistically narrow, leading to biased contamination estimates from the multicolour
analysis.

\begin{figure}
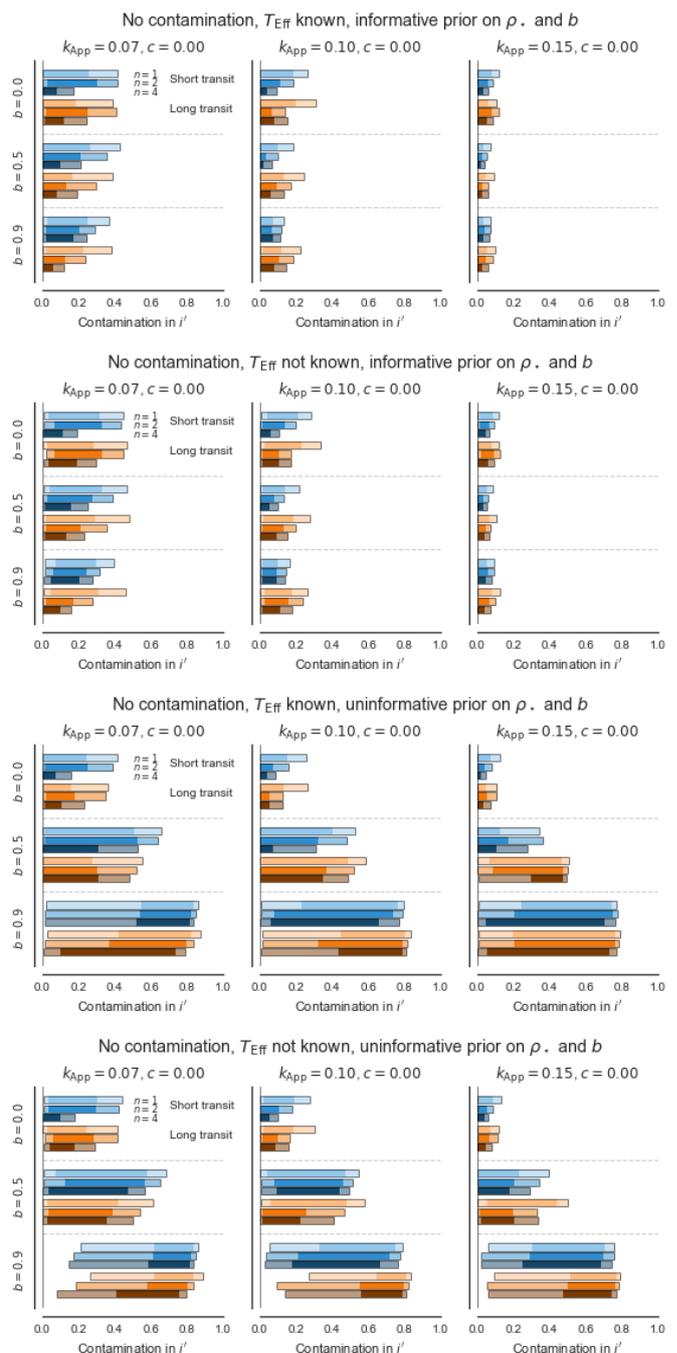

	\centering
	\includegraphics[width=\linewidth]{sim_cn_MGi_00_constrained}
	\includegraphics[width=\linewidth]{sim_cn_MGu_00_constrained}
		\includegraphics[width=\linewidth]{sim_cn_MGi_00}
	\includegraphics[width=\linewidth]{sim_cn_MGu_00}
	\caption{The $50\%$ and $95\%$ posterior limits (darker and lighter colouring, respectively) for the estimated contamination in the $i'$ band
		for an uncontaminated M dwarf with either informative or uninformative priors on \prho and \pimp, and either informative
		or uninformative prior on \teffh. The results are shown for three impact parameters (\pimp), 
		three apparent radius ratios in the $i'$ band (\para) , three dataset sizes ($n$, number of observed transits), and two transit durations,
		as detailed in Sect.~\ref{sec:simulations.overview}.
		Blue colour corresponds to short (1~h) transit duration, and orange to long (2~h) transit duration.}
	\label{fig:sim_mgi_00}
\end{figure}

Figure~\ref{fig:sim_mgi_00} collects the results from the uncontaminated simulations.
The first two sub-figures (from the top) show the simulations with an informative prior on 
the orbital parameters, and the last two without (basically, the amount of prior information
decreases from top to bottom). The columns separate the three different apparent radius ratios, and
each sub-figure is divided vertically by the impact parameter, and further by the number 
of transits observed.

When the orbit is constrained by a prior, the marginal reference passband contamination (\pcref)
posteriors are also well-constrained and have their modes at (or close-to) $0$ for all
impact parameters, no matter whether we set a prior on \teffh or not. This shows that significant
contamination, when there is none, can be ruled out using a relatively small amount of
ground-based photometry if we have reliable constraints on the orbital parameters.

When the orbit is unconstrained, the marginal contamination becomes less constrained as
the impact parameter increases, and the posterior mode moves towards higher values of
contamination. This effect becomes increasingly prominent when we set an uninformative
prior on \teffh.

\begin{figure*}
	\centering
	\includegraphics[width=\linewidth]{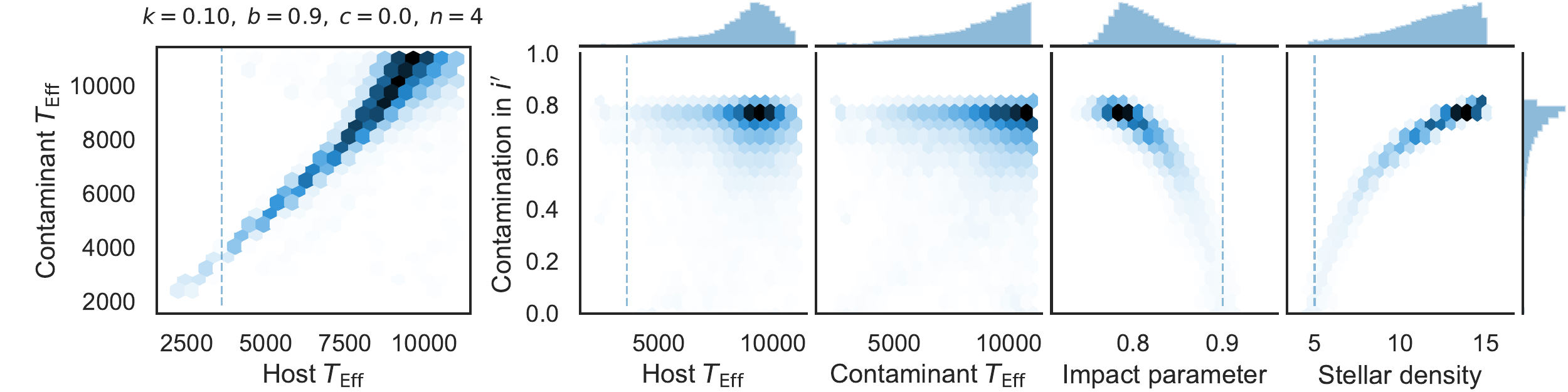}
	\includegraphics[width=\linewidth]{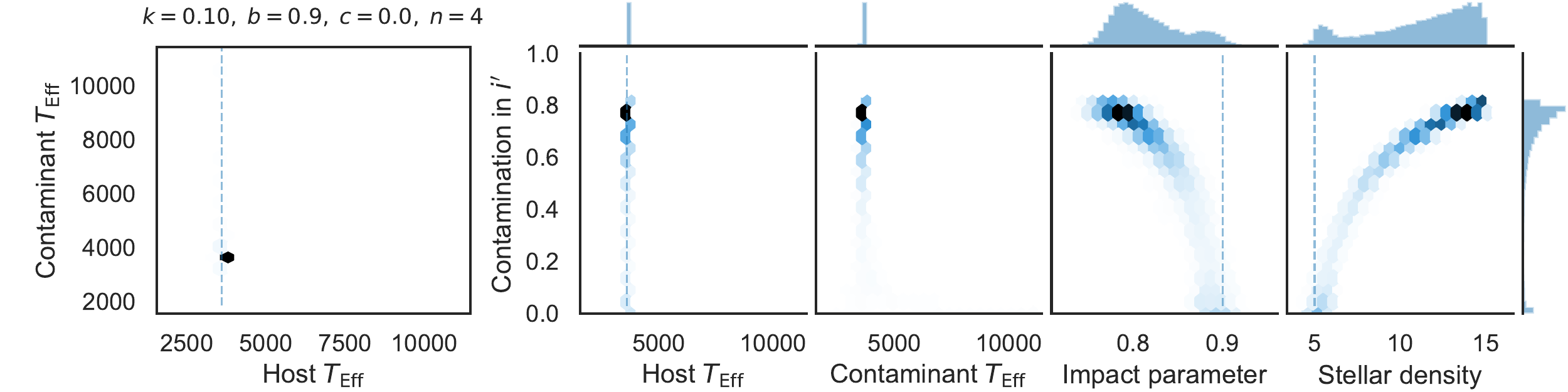}
	\includegraphics[width=\linewidth]{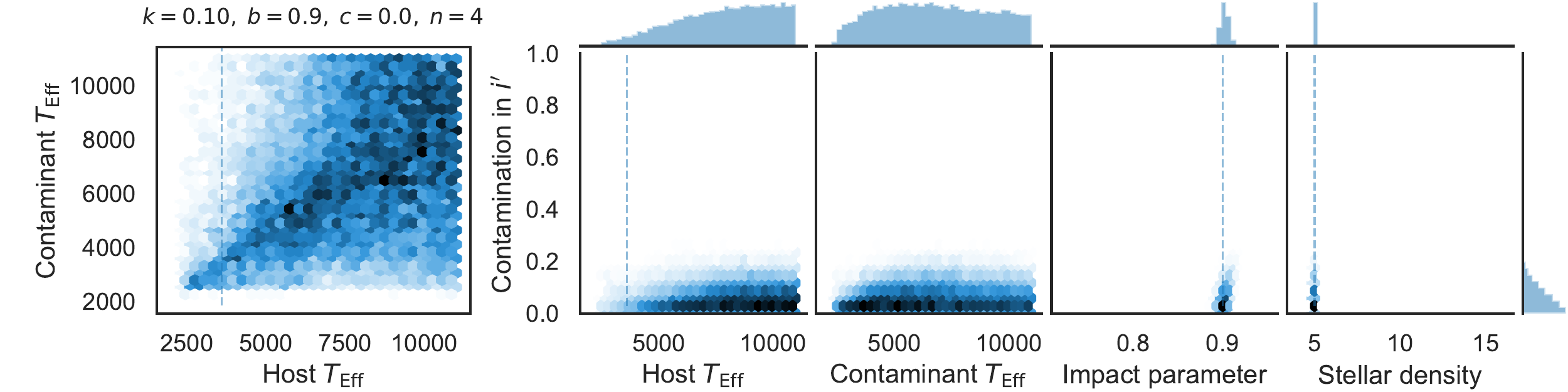}
	\includegraphics[width=\linewidth]{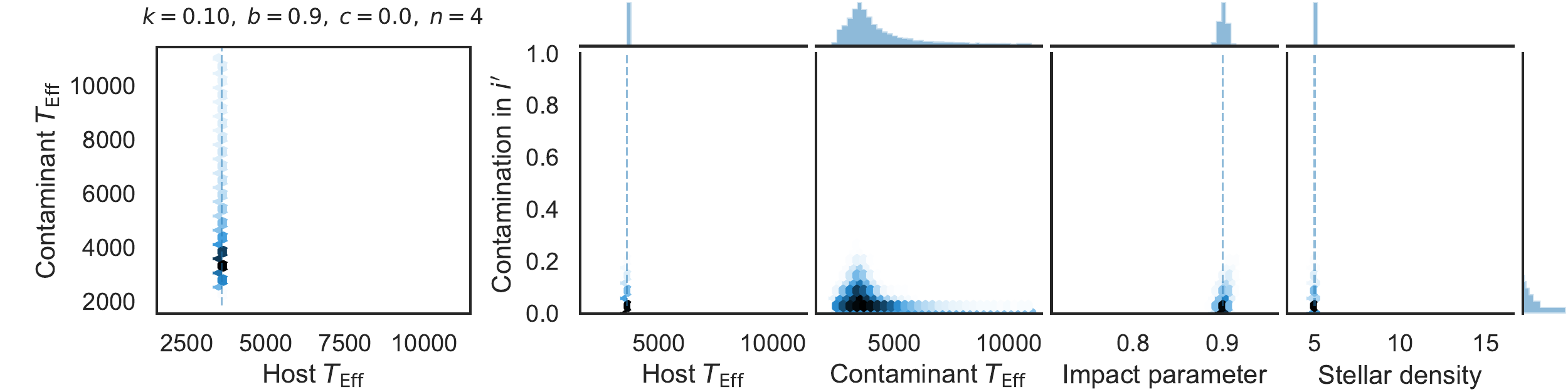}
	\caption{Joint posterior distributions for \teffh, \teffc, \pimp, \prho, and \pcref for the case without contamination,
		$\para = 0.1$, and $\pimp = 0.9$. In this example the high impact parameter leads to weakly constrained $(\pcref,\; \pimp,\; \prho)$-space.
	The 1st  row shows the distributions without external constraints, the 2nd row with a prior on \teffh, the 3rd row with priors on \pimp and \prho,
	and the 4th row with priors on \teffh,  $b$, and \prho.}
	\label{fig:sim_m00_joints}
\end{figure*}

When we look at the \pcref, \teffh, \teffc, \pimp, and \prho joint distributions, shown in
Fig.~\ref{fig:sim_m00_joints} for $\para=0.1$, $\pimp=0.9$, we find patterns that help us to
understand what is happening. With the orbit and \teffh unconstrained (1st row in
Fig.~\ref{fig:sim_m00_joints}), the stellar density, impact parameter, and contamination
are increasingly degenerate for higher impact parameter values, and orbits with low impact
parameter, high stellar density, and high contamination are favoured when the orbit is
constrained only by the ground-based photometry. Constraining the host \teff (2nd row in
Fig.~\ref{fig:sim_m00_joints}) reduces the degeneracy and leads to bimodal marginal \pimp,
\prho, and \pcref distributions, where the weaker mode corresponds to the true
configuration. When we set tight priors on \pimp and \prho, but assume no information
about \teffh (3rd row in Fig.~\ref{fig:sim_m00_joints}), the degeneracy between the
orbital parameters and contamination disappears. When we add constraints on \teffh (4th row
in Fig.~\ref{fig:sim_m00_joints}), the allowed contamination space decreases further. Now
the maximum amount of contamination is constrained strongly, and any significant differences
between \teffc and \teffh are rejected.
 
Even in the least-constrained case the joint distributions yield useful information that
can be used together with other planet candidate vetting methods. First, the \teffc and
\teffh distribution tells that if the photometry is contaminated, the contamination is
from source(s) with a similar \teff as the host (or that the combined spectrum from
multiple sources matches closely the host star's spectrum). Further, the stellar density
distribution can be used with external information to assess how likely the host star is
to be of the predicted density.

\subsection{M-dwarf slightly contaminated by a G star}
\label{sec:simulations.mg}

Next, we study the same M dwarf as in Sect.~\ref{sec:simulations.m_no_contamination}, but
now slightly contaminated (10\% of the total flux) by a $\teff=5800$~K G star. This should
be an easy-to-identify scenario since the colour difference between the two stars leads to significant
transit depth variations across passbands. Again, we consider the effect of setting
informative priors on \teffh and the orbital parameters, and we assume that the prior on
\teffh corresponds to the true \teffh since the host dominates the total flux.

Figure~\ref{fig:sim_mgi_01} collects the results from the low-contamination simulations
with host-contaminant colour difference. Now, the most constrained case (where we set
informative priors on \teffh, \pimp, and \prho) yields reliable contamination estimates
for all the simulation scenarios, whereas the least constrained case  still yields
reliable estimates for the high signal-to-noise transit light curves with centric transits. 
Having information about the orbital parameters allows us to again estimate the
high-impact-parameter candidates, but does not significantly affect the
low-impact-parameter estimates. Finally, setting a prior on \teffh significantly improves
the contamination estimates for all cases, even for $\pimp=0.9$. Specially, 
unconstrained \teffh with low signal to noise leads to a bias towards high contamination
values. This can be explained by the nonlinear relation between the contamination, apparent
radius ratio, and true radius ratio. Contamination scales as $c = 1 - \paaa / \ptaa$, so
a small increase in the range of values the true radius ratio can have can lead to a significant
increase in the contamination values (see Appendix~\ref{sec:appendix.model} for details).

In most cases, having reliable information about \teffh is more important than being able
to constrain the orbit. The \teffh estimate can be obtained either from spectroscopy, or
roughly from the multicolour photometry itself. However, this estimate may be
erroneous if the contaminant dominates the total observed flux. This case is the topic of
the next section.

\begin{figure}
	\centering
	\includegraphics[width=\linewidth]{sim_cn_MGi_co_01}
	\includegraphics[width=\linewidth]{sim_cn_MGu_co_01}
		\includegraphics[width=\linewidth]{sim_cn_MGi_01}
	\includegraphics[width=\linewidth]{sim_cn_MGu_01}
	\caption{The $50\%$ and $95\%$
		posterior limits for the estimated contamination in the $i'$ band for an M dwarf slightly
		contaminated by a G2 star (10\% of the total flux) with or without an informative prior
		on $\rho_\star$ and $b$ (first two rows from the top and the last two rows from the top,
		respectively), and either informative or uninformative prior on \teffh (first and third,
		and second and fourth rows, respectively ). The results are shown for three impact
		parameters (\pimp), three apparent radius ratios in the $i'$ band (\para) , three dataset
		sizes ($n$), and two transit durations, as detailed in
		Sect.~\ref{sec:simulations.overview}. Blue colour corresponds to short (1~h) transit
		duration, and orange to long (2~h) transit duration.}
	\label{fig:sim_mgi_01}
\end{figure}

\subsection{M-dwarf heavily contaminated by a G star}
\label{sec:simulations.strong_colour_difference}

Considering that even a slight contamination by a star with significantly different \teff
is relatively easy to measure, we can expect the detection of heavy contamination to be
easy as well. Now, the main issue of interest is what happens if we mistake the
contaminant as the candidate host star, and set an informative prior on \teffh that
actually corresponds to \teffc.

\begin{figure*}
	\centering
	\includegraphics[width=\linewidth]{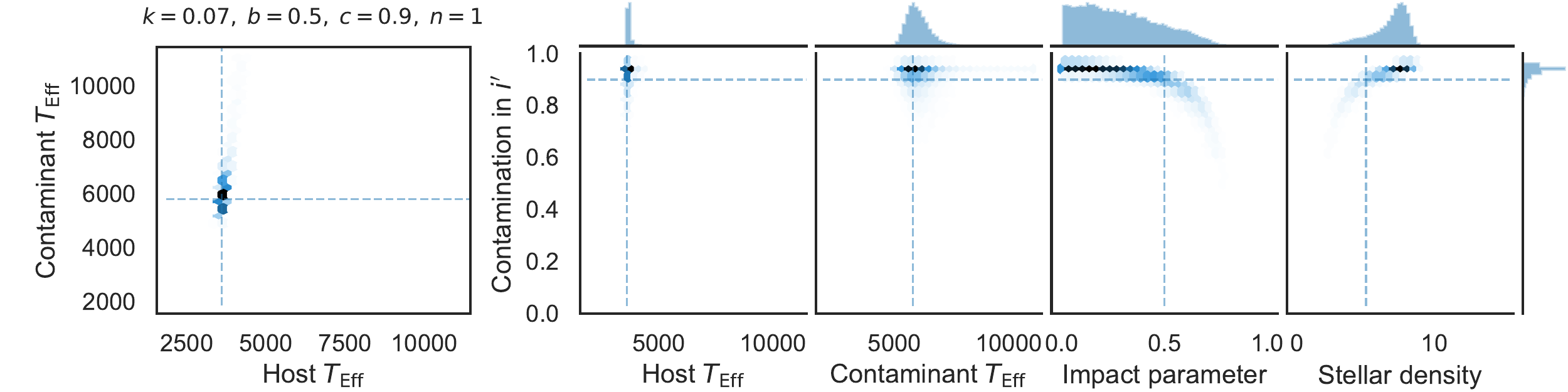}
	\caption{As in Fig.~\ref{fig:sim_m00_joints}, but for an M dwarf heavily contaminated by a G star ($\pcref=0.9$), and with uninformative priors on \teffh and orbital parameters. The colour difference
	between the stars leads to strong transit depth variations and allows us to determine \teffh, \teffc,
	and \pcref accurately.}
	\label{fig:sim_mg_ftfo_90_joint}
\end{figure*}

Figure~\ref{fig:sim_mgi_90} collects the results from the high-contamination ($\pcref =
0.9$) simulations with host-contaminant colour difference, and
Fig.~\ref{fig:sim_mg_ftfo_90_joint} shows the joint posteriors for a single simulation
case with short (1~h) transit duration, uninformative priors on \teffh and orbital
parameters, $\para = 0.07$, and $\pimp = 0.5$ for one observed transit.

As expected, we recover accurate \teffh, \teffc, and contamination estimates in the case of
uninformative prior on \teffh, even with low SN observations. Surprisingly, setting a
(wrong) informative prior on \teffh has a relatively small effect on the final
contamination estimate. This can be explained by degeneracies in the relative fluxes in
the (\teffc, \teffh)-space. Even when forcing \teffh close to 5800~K, we can discover a
solution with $\teffc\sim12\,000$~K (at the boundary of the contamination model space)
that sufficiently explains the transit depth differences  (that is, the relative colour differences
between the two cases are similar enough to yield a reliable contamination estimate.)
However, this result cannot be generalised, since the situation will likely be different
for a hotter host star.

Thus, as expected, significant contamination from a contaminant with $\teffc \ll \teffh$
or $\teffc \gg \teffh$ can be estimated reliably without imposing an informative prior on
\teffh due to the transit depth differences.

\begin{figure}
	\centering \includegraphics[width=\linewidth]{sim_cn_MG_ctco_09}
	\includegraphics[width=\linewidth]{sim_cn_MG_ftco_09}
	\includegraphics[width=\linewidth]{sim_cn_MG_ctfo_09}
	\includegraphics[width=\linewidth]{sim_cn_MG_ftfo_09} \caption{The $50\%$ and $95\%$
	posterior limits for the estimated contamination in the $i'$ band for an M dwarf strongly
	contaminated by a G2 star (90\% of the total flux) with or without an informative prior
	on $\rho_\star$ and $b$ (first two rows from the top and the last two rows from the top,
	respectively), and either informative or uninformative prior on \teffh (first and third,
	and second and fourth rows, respectively ). The results are shown for three impact
	parameters (\pimp), three apparent radius ratios in the $i'$ band (\para) , three dataset
	sizes ($n$), and two transit durations, as detailed in
	Sect.~\ref{sec:simulations.overview}. Blue colour corresponds to short (1~h) transit
	duration, and orange to long (2~h) transit duration.} 
\label{fig:sim_mgi_90}
\end{figure}

\subsection{M-dwarf contaminated by an M-dwarf}

We next focus our attention to the case of contamination when $\teffc \sim \teffh$. Now
the lack of colour difference means that contamination does not lead to transit depth
variations across passbands, and all of the information about it is in
the chromatic variations in the curvature of the transit light curve.

We change the simulation setup slightly, and study contamination factors $\pcref = \{0.25,\, 0.50,\, 0.85\}$, 
since the $\pcref = 0.0$ case was already studied in Sec.~\ref{sec:simulations.m_no_contamination}.

\begin{figure*}
	\centering 
	\includegraphics[width=0.49\linewidth]{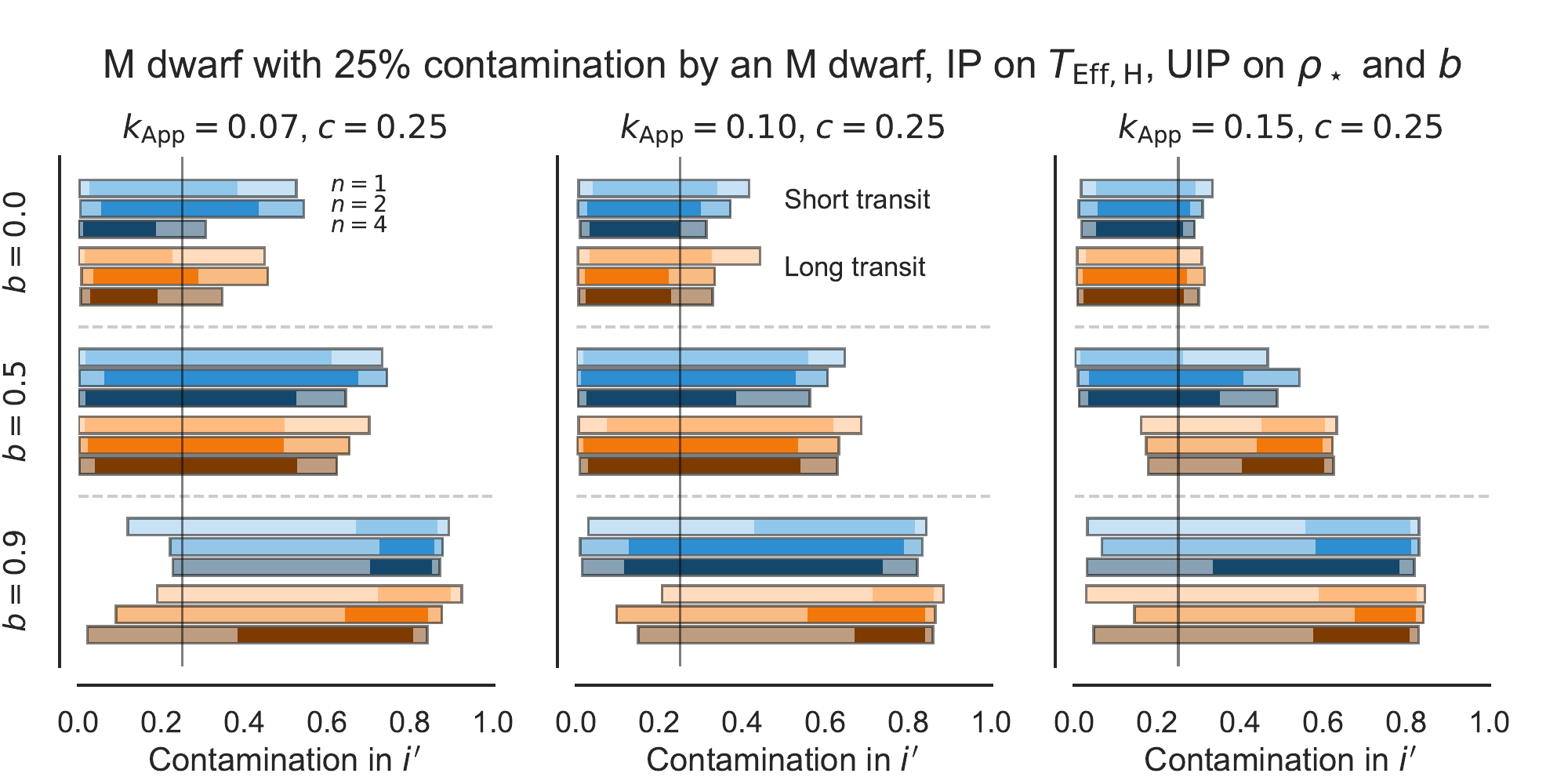}
	\includegraphics[width=0.49\linewidth]{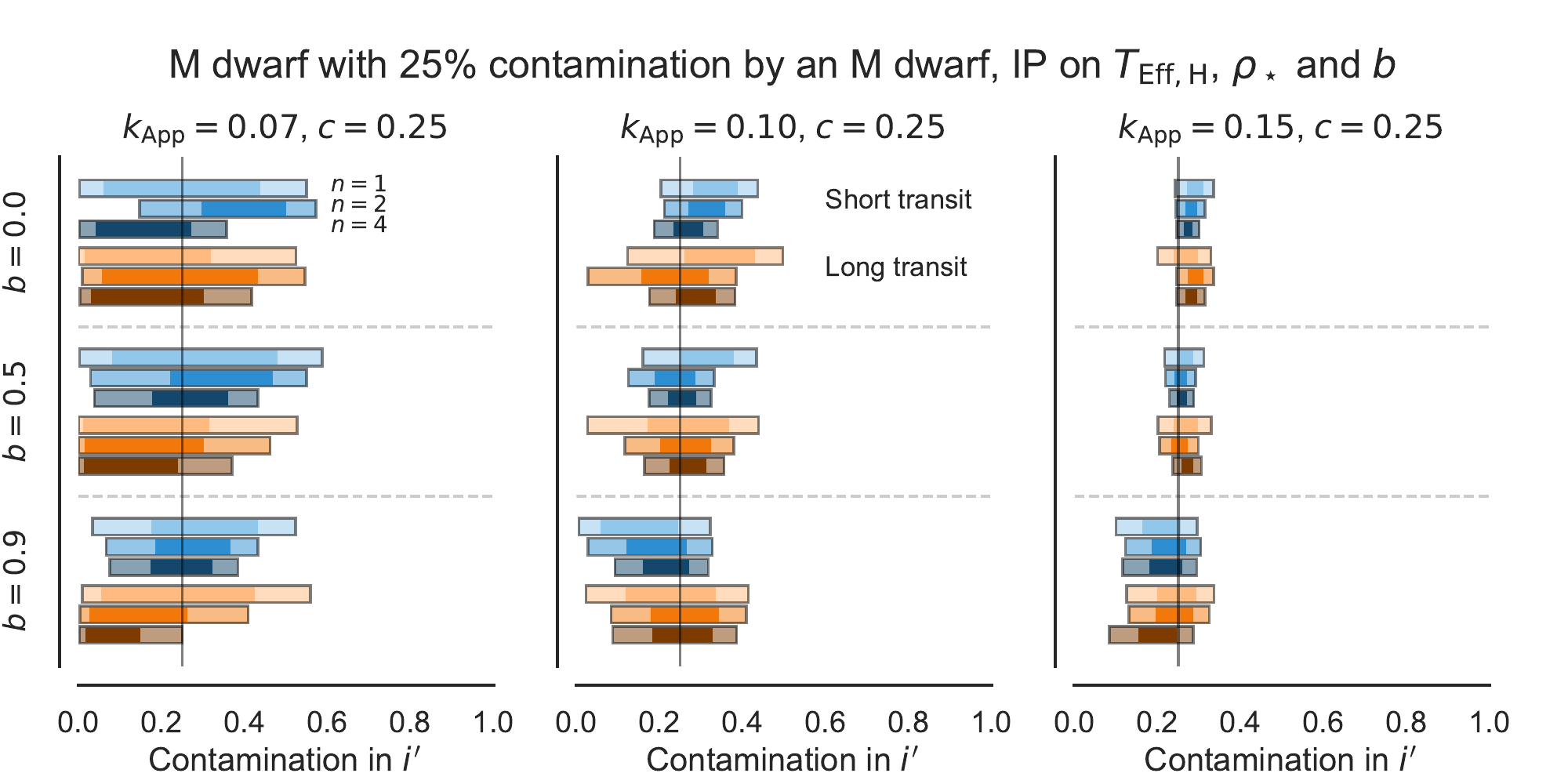}\\
	\includegraphics[width=0.49\linewidth]{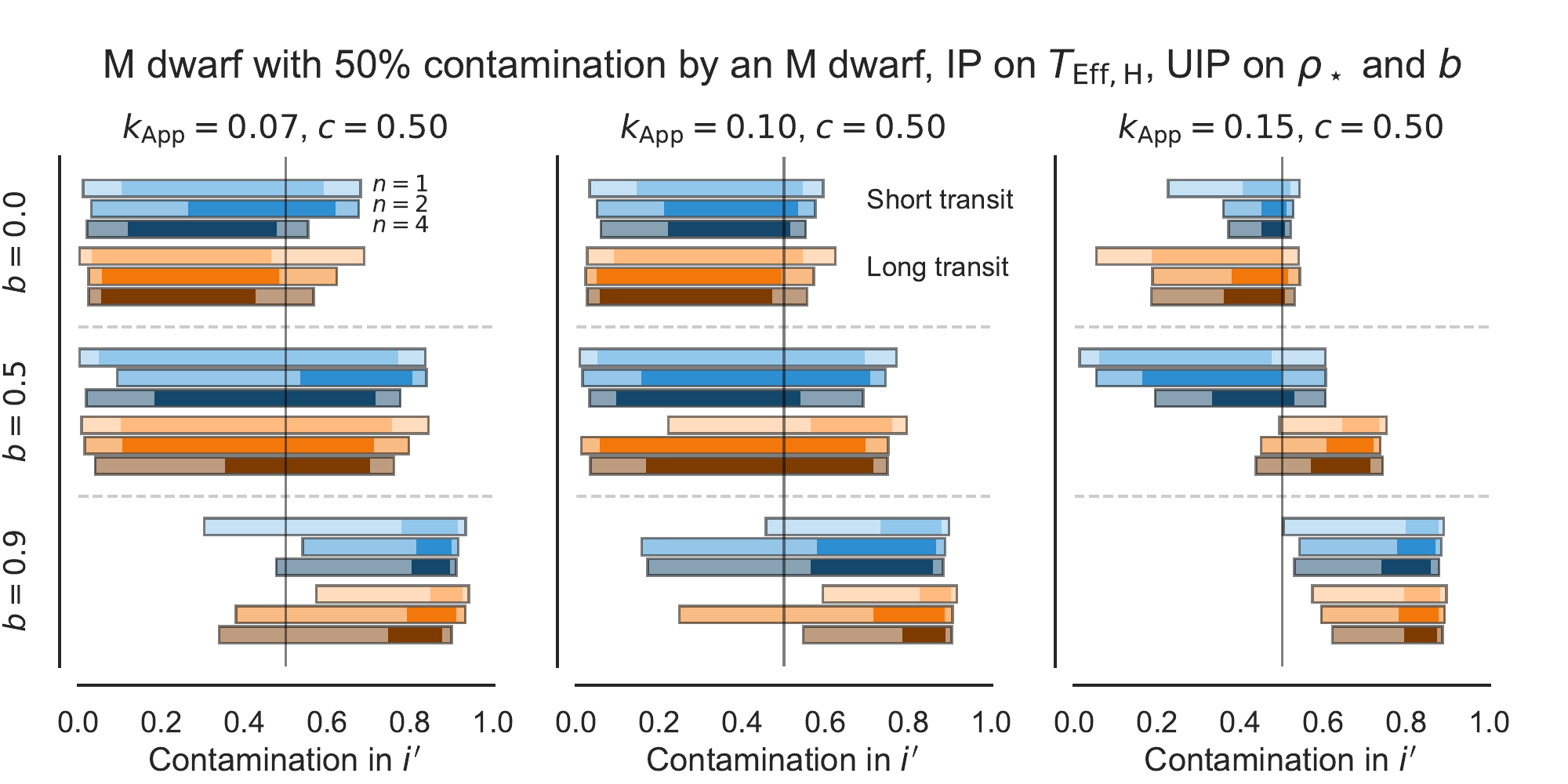}
	\includegraphics[width=0.49\linewidth]{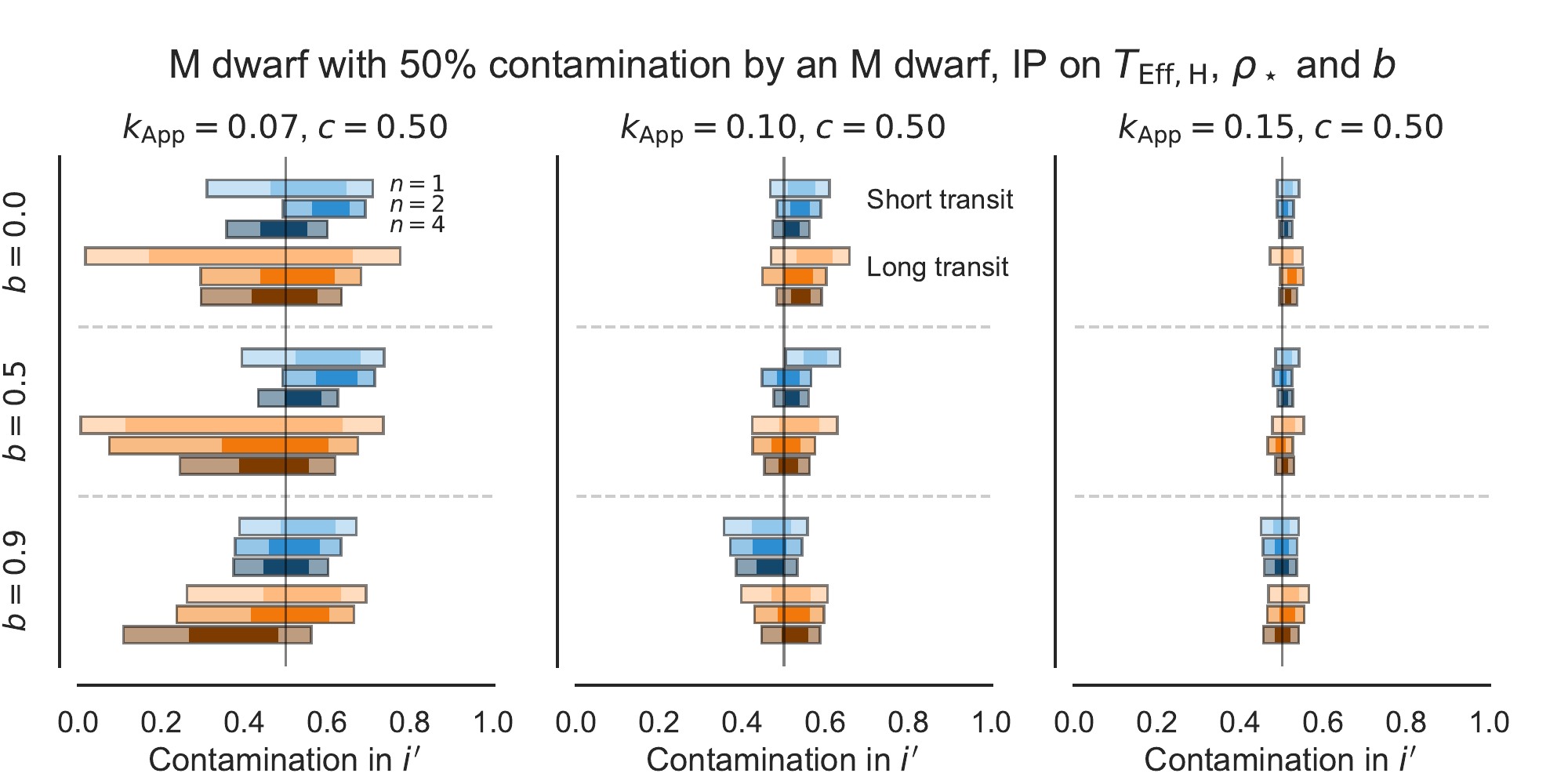}\\
	\includegraphics[width=0.49\linewidth]{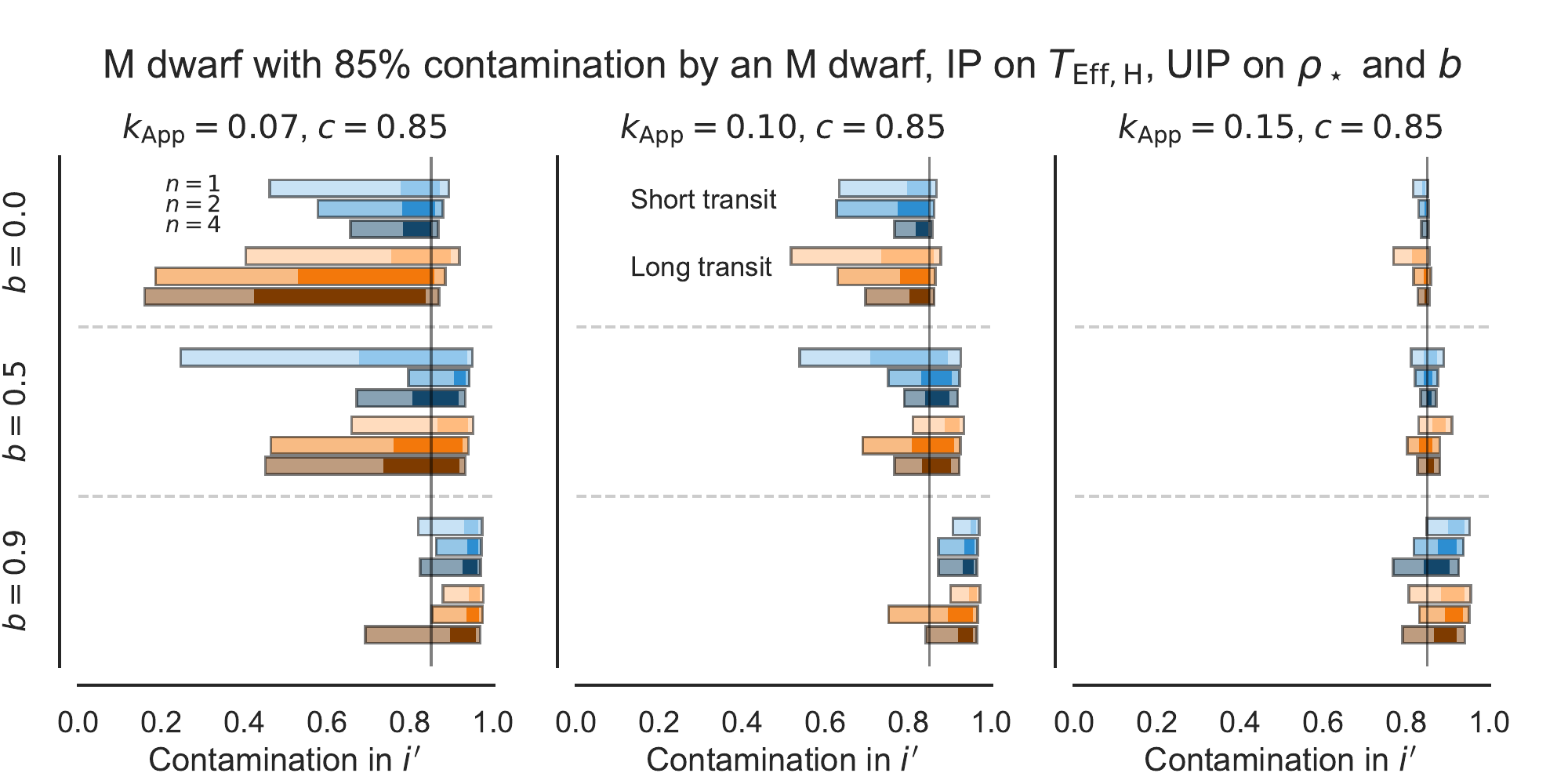}
	\includegraphics[width=0.49\linewidth]{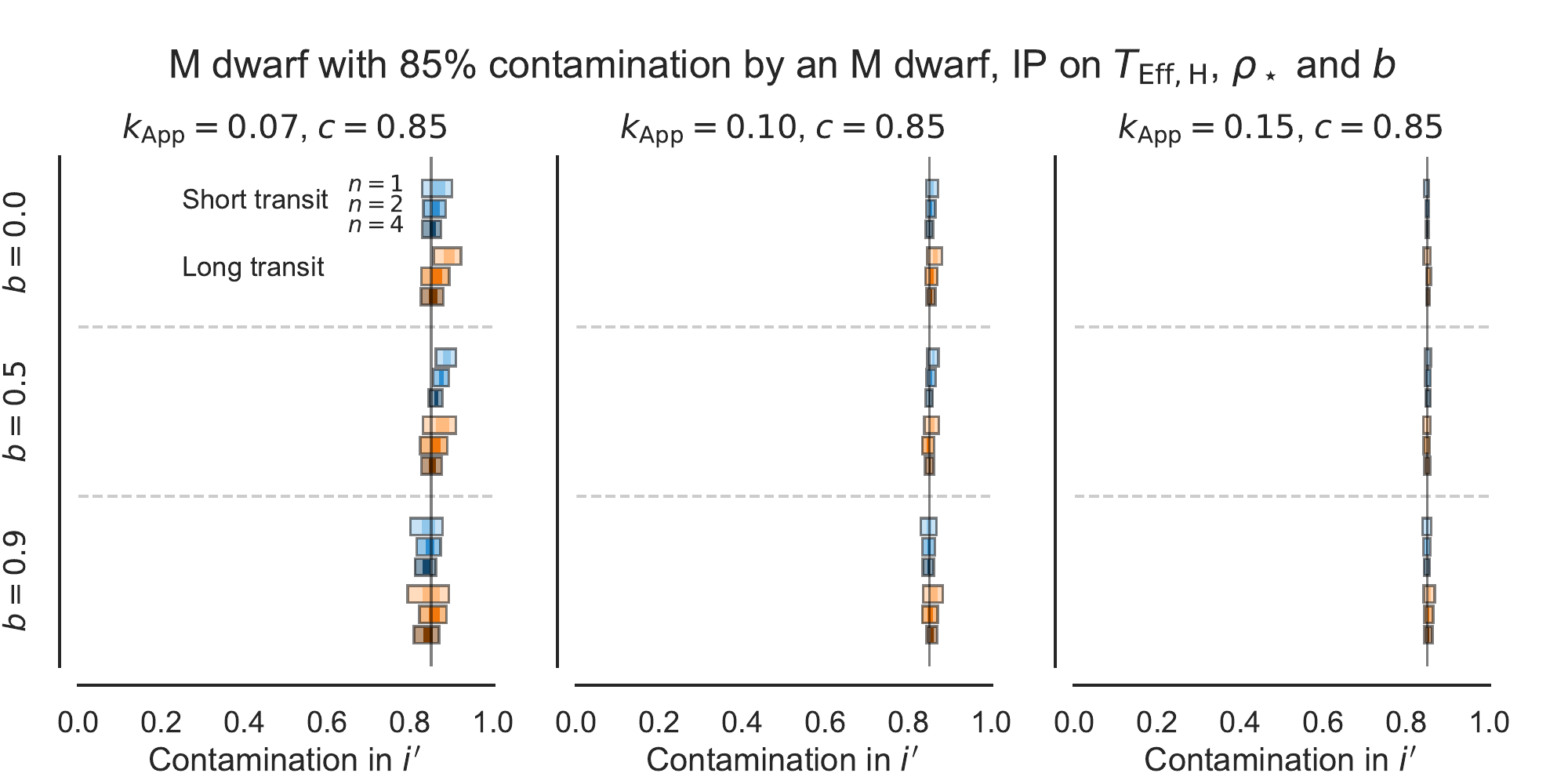}
	\caption{The $50\%$ and $95\%$
		posterior limits for the estimated contamination in the $i'$ band for an M dwarf 
		contaminated by an identical M dwarf with an informative prior on \teffh (see the text why this has no impact),
		with or without an informative prior
		on $\rho_\star$ and $b$ (left and right columns, respectively), and 25\%, 50\%, and 85\% of
		contamination in the $i'$ band, from top to bottom. The results are shown for three impact
		parameters (\pimp), three apparent radius ratios in the $i'$ band (\para) , three dataset
		sizes ($n$), and two transit durations, as detailed in
		Sect.~\ref{sec:simulations.overview}. Blue colour corresponds to short (1~h) transit
		duration, and orange to long (2~h) transit duration.} 
	\label{fig:sim_mm}
\end{figure*}

We show the result in Fig.~\ref{fig:sim_mm} for three contamination levels, informative
prior on \teffh, and either with or without informative priors on \pimp and \prho.
Constraining \teffh does not affect parameter posteriors other than \teffh and \teffc,
since the second colour signature does not depend on the absolute \teffh (we carried out
the simulations for cases without an informative prior on \teffh to ensure this is the
case). Thus, In the following we focus only on the cases with an informative prior on
\teffh.

Because of the degeneracies, except for the largest planets and highest contamination levels, the contamination is
poorly constrained when we do not have informative priors on the orbital parameters.
However, constraining \pimp and \prho constrains the contamination estimates significantly. 

Further constraints could be obtained by constraining the stellar limb darkening. This however
goes outside the scope of the simulation study, but is done in the analysis of real observations
in Sect.~\ref{sec:reality}.

\subsection{Scaling to small planets and low signal to noise}

We carry out a separate set of simulations to study how the approach extends to small planet
candidates observed with large telescopes. We choose $\para = 0.02$ corresponding
to an Earth-sized planet around an M-star and run the simulations for $\pcref = \{0.00, 0.99\}$
without host-contaminant colour difference, and for $\pcref = 0.99$ with an
G2 contaminant, as before. The true radius ratio, \ptra, for the contaminated case ($\pcref = 0.99$) is now 0.2. 
We repeat the simulations for two white noise levels, 100~ppm and 200~ppm
over an exposure of 1 minute, and three sets of observed transits ($n_\mathrm{n} = \{1, 2, 4\}$),
but restrict to $\pimp=0$, a short transit duration with $T_{14} \sim 1$~h, and uninformative priors 
on \teffc, \prho, and \pimp. We do not show the results here, but they are available from
the paper's GitHub repository.

All the high-contamination cases are identified correctly, even without host-contaminant colour difference. 

However, the contamination estimates for the no-contamination cases are biased towards high contamination 
values. This is because of the low SN ratio and the nonlinear relationship between the contamination,
apparent radius ratio, and true radius ratio. Contamination from sources with colour difference is
still ruled out, but the transit shape cannot constrain contamination from sources without a
colour difference.

More than four transits are needed to rule out significant contamination from a source with $\teffc = \teffh$ 
when $\sigma = 200$~ppm over a one minute exposure. However, a single transit is sufficient with 
$\sigma = 100$~ppm to ensure that the true radius ratio is no more than 2-3 times the apparent radius
ratio, which is sufficient to validate a super-Earth (this corresponds to maximum contamination of 0.9, 
see Appendix~\ref{sec:appendix.model} for details).

While the required precision is high, a it may achievable from the ground with the currently
existing large telescopes, such as the 10.3~m Gran Telescopio Canarias (GTC), and possibly also with the smaller 
telescopes with the aid of a photometric diffuser \citep{Stefansson2017,VonEssen2019}.

\section{Application to real data}
\label{sec:reality}
\subsection{MuSCAT2 Observations}

We use transit observations of WASP-43b and WASP-12b to study the
contamination estimation in practice. The observations were carried out with \textit{MuSCAT2}, a
new four-colour imager installed in the 1.5~m Carlos Sanchez Telescope (TCS) in the Teide
Observatory (OT), Tenerife \citep{Narita2018}. The instrument has
the capability of simultaneous imaging in $g'$ (400-550~nm), $r'$ (550-700~nm), $i'$
(700-820~nm), and $z_s$ (820-920~nm) bands, which closely match our simulations.

\subsection{Data reduction and analysis}

The photometry was carried out with a dedicated MuSCAT2 photometry pipeline based on \textsc{NumPy}, 
\textsc{SciPy}, \textsc{AstroPy} \citep{TheAstropyCollaboration2013}, \textsc{photutils} \citep{Bradley2019},
and \textsc{astrometry.net} \citep{Lang2010}. The contamination analysis was carried out using MuSCAT2 
transit analysis pipeline relying on \pytransit, \textsc{george}, \textsc{emcee}, and \textsc{PyDE}.

The light curve modelling was carried out with a MuSCAT2 transit analysis pipeline that
uses \pytransit~v2 for blending simulation and \ldtk to constrain limb darkening. We
first modelled each night separately to test for possible night-to-night variations in any of the
parameters of interest, and then carried out a joint modelling using all the nights
simultaneously. The analysis starts by fitting a transit to the data with a linear
baseline model. Next, we repeat the analysis with a Gaussian process (GP) -based systematics model where the
GP hyperparameters are fixed to values optimised to the light curves with the best-fitting
linear-baseline-model transit removed. The initial fitting is carried out using a Differential
Evolution global optimisation method that results with a population of parameter vectors
clumped close to the global posterior mode. We use this parameter vector population as a
starting population for the MCMC sampling with \textit{emcee}, and carry out the sampling
until the we obtain a reliable posterior sample \citep{Parviainen2018}.

The MCMC sampling was repeated both for the linear and GP systematics models to test for
possible differences in the inference due to the different systematics modelling
approaches. The posteriors from the two approaches agreed with each other, and we adopted
the GP-based posteriors for the rest of the study.

\subsection{WASP-43b}

WASP-43b is a short-period hot Jupiter transiting a $\teff = 4400~K$ K7V star every 0.81~d \citep{Hellier2011,Gillon2012}. The planet
is well-studied \citep{Murgas2014,Chen2014}, and is as a good example case of an easy target for contamination
estimation.

\begin{figure*}
	\centering
	\includegraphics[width=\linewidth]{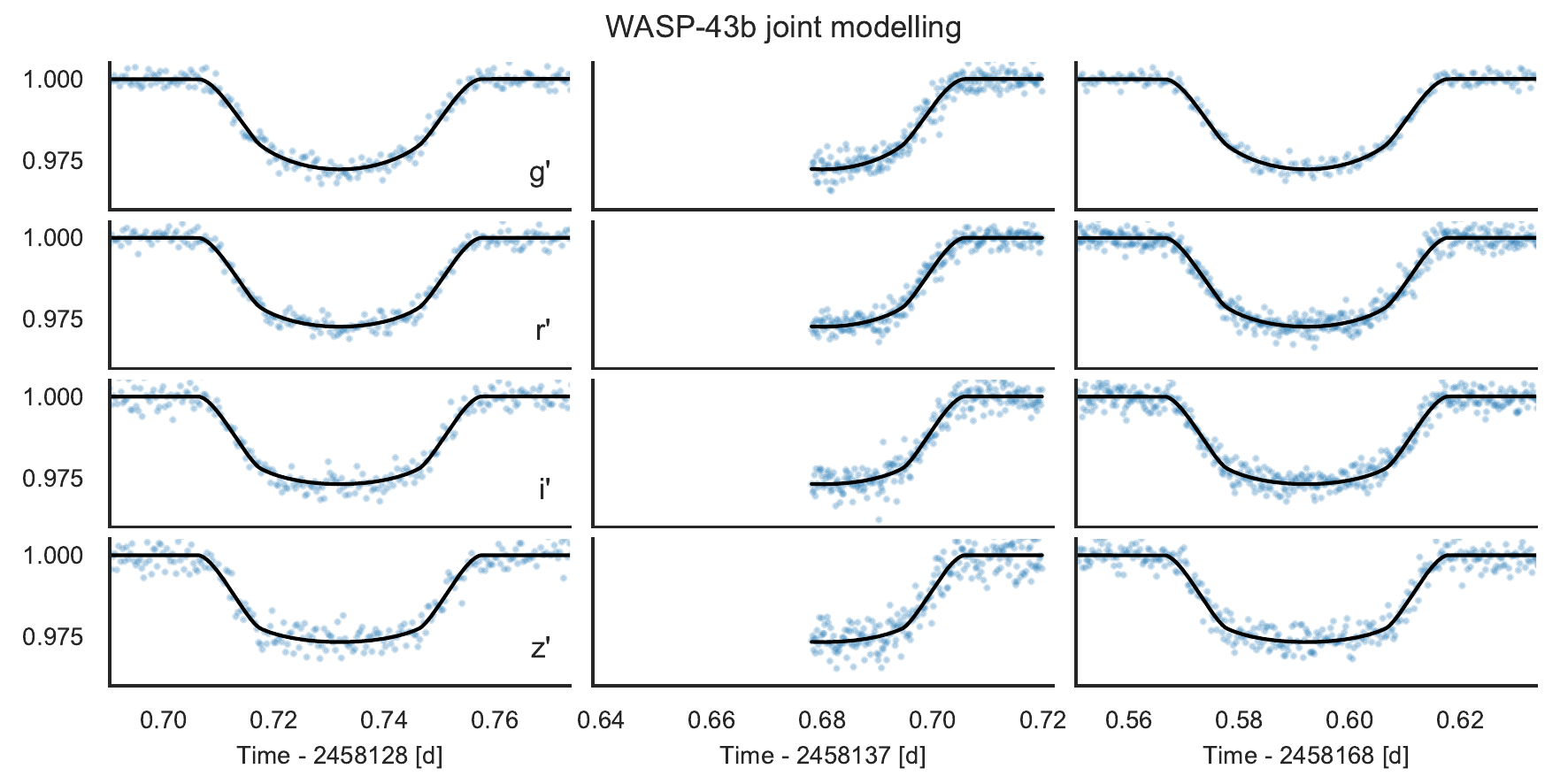}
	\caption{Two full and one partial transits of WASP-43b observed with MuSCAT2 in $g'$, $r'$, $i'$, and $z'$ with a
		transit model corresponding to the median of the model parameter posteriors. The systematics have been modelled
		using a Gaussian process, and the GP mean has been removed from the observed data for visualization purposes.}
	\label{fig:wasp43b_lcs}
\end{figure*}

\begin{figure*}
	\centering
	\includegraphics[width=\linewidth]{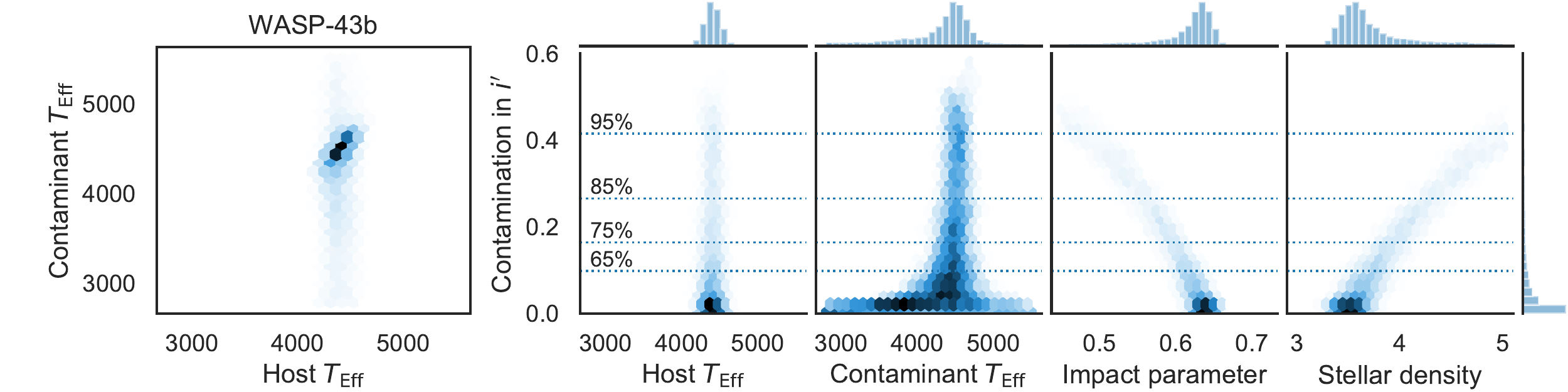}
	\includegraphics[width=\linewidth]{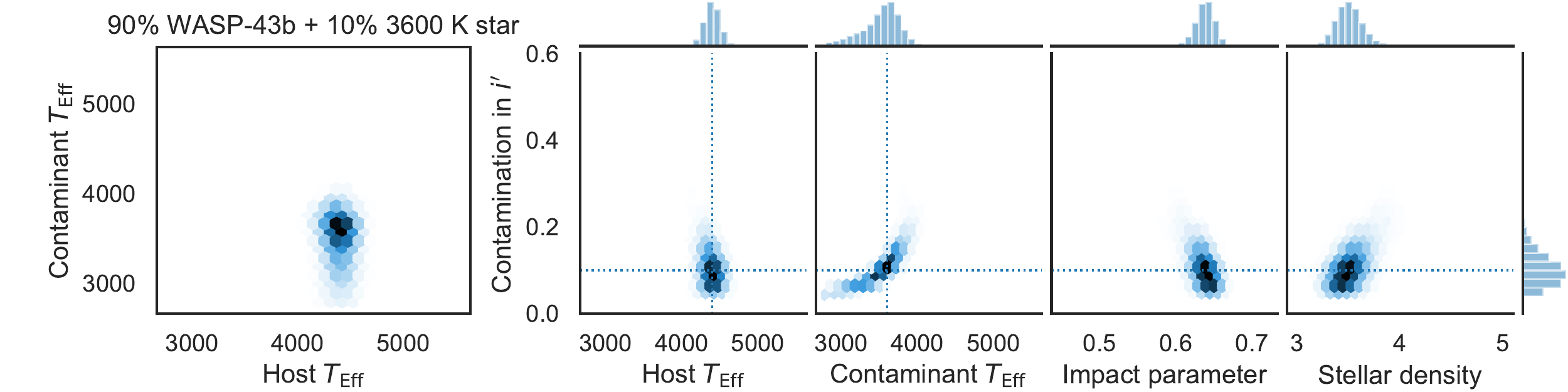}
	\includegraphics[width=\linewidth]{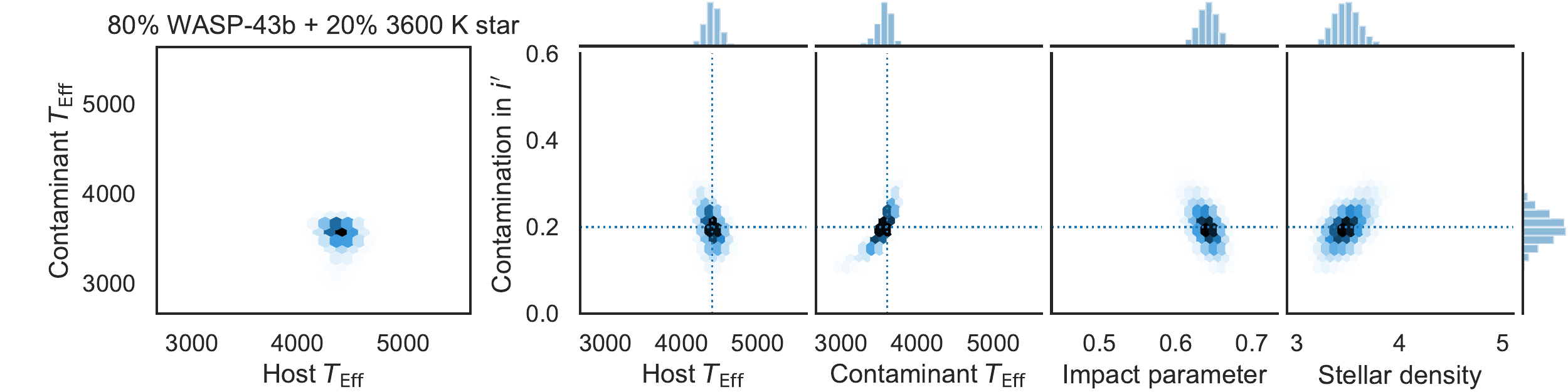}
	\caption{WASP-43b blending analysis using the MuSCAT2-observed light curves without injected contamination (above), with
		10\% contamination from a 3600~K star (middle), and with 20\% contamination from a 3600~K star (bottom). The dotted lines
	shows the 65\%, 75\%, 85\%, and 95\% percentiles for the posterior contamination.}
	\label{fig:wasp43b_posteriors}
\end{figure*}

The observations cover two full and one partial transit of WASP-43b on nights 9.1.2018, 18.1.2018, and 18.2.2018, shown in
Fig.~\ref{fig:wasp43b_lcs}. We first carry out the analysis for original observations, and then repeat the analysis for cases
with 10\% and 20\% of injected contamination from a 3600~K star. The results are shown in  Fig.~\ref{fig:wasp43b_posteriors}.
 
In the first case (no injected colour-dependent contamination) the observations allow us
to reject any significant differences between the host and a possible contaminant effective temperatures.
The contamination is also constrained to relatively low values, and could be constrained
further by setting a prior on stellar density. In the two cases with small levels of injected
contamination from a 3600~K star, the contaminant properties and the level of
contamination are retrieved faithfully.

\subsection{WASP-12b}
\label{sec:reality.wasp12b}

WASP-12b is an inflated hot Jupiter orbiting the  $\teff= 6300$~K primary of a hierarchical triple star
system  \citep{Bechter2013}. The existence of two close-in M-star companions was originally
missed \citep{Hebb2009}, and the two companion stars were first identified as a single M star
by \citet{Crossfield2012}  with T$_\mathrm{Eff} = 3660 \pm 70$~K, and combined primary-companion
flux ratios of 0.08 in the $H$ band, and 0.03 in the $i'$ band. The presence of
well-characterised contaminating stars makes WASP-12b an interesting target to test the
contamination estimation in practise. The two companions are of same type, which allows us
to carry out a simple simulation with one contaminant.

\begin{figure*}
	\centering
	\includegraphics[width=\textwidth]{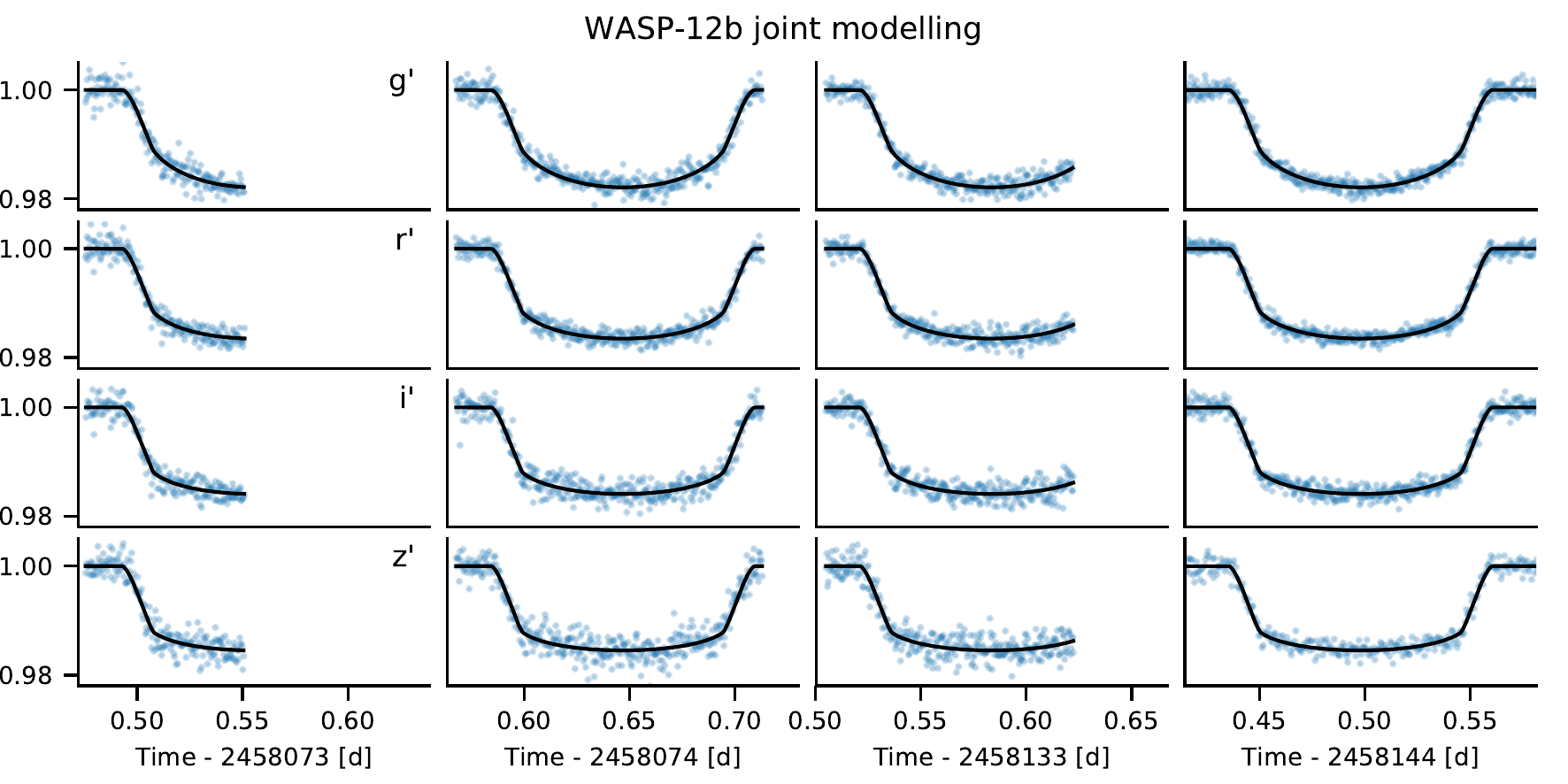}
	\caption{Two full and two partial transits of WASP-12b observed with MuSCAT2 in $g'$, $r'$, $i'$, and $z'$. The light curves have been detrended
		using a GP with a conservative kernel for visualization purposes.}
	\label{fig:wasp12b_lcs}
\end{figure*}

\begin{figure*}
	\centering
	\includegraphics[width=\linewidth]{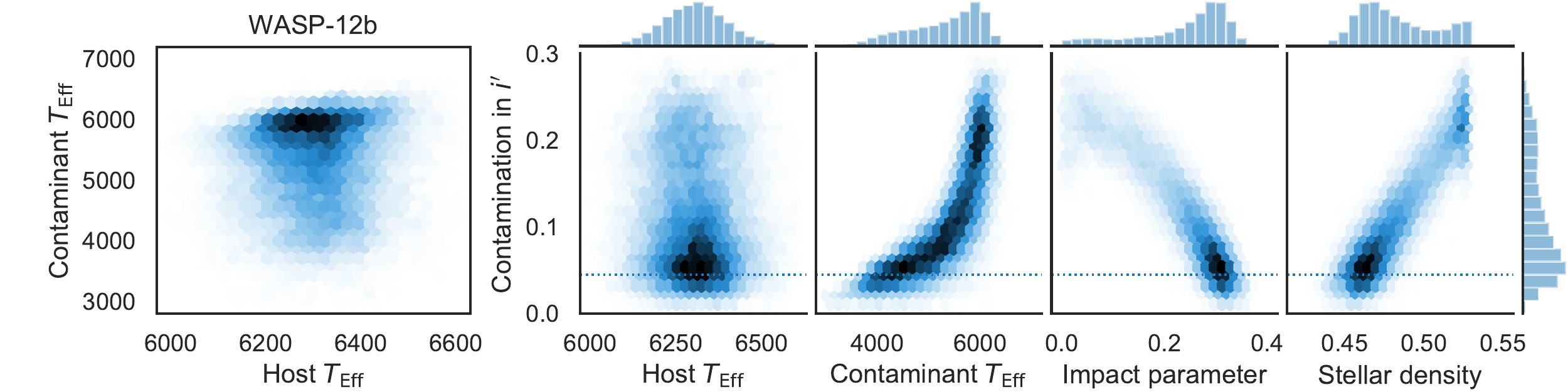}
	\caption{WASP-12b blending analysis using the MuSCAT2-observed light curves  shown in Fig.~\ref{fig:wasp12b_lcs}.
	WASP-12 is a hierarchical triple system with the planet transiting a $\teff=6300$~K  primary. The photometry
	is contaminated by two unresolved M-star companions contributing $\sim4\%$ of flux in the $i'$ band.}
	\label{fig:wasp12b-posteriors}
\end{figure*}

We observed two full and two partial four-colour transits of WASP-12b with MuSCAT2 on
nights 15.11.2017, 16.11.2017, 14.1.2018, and 25.1.2018, shown in
Fig.~\ref{fig:wasp12b_lcs}.  \ldtk was used to constrain the limb darkening assuming
$\teff = 6300 \pm 50$~K, $\log g = 4.38 \pm 0.1$, and $Z = 0.3 \pm 0.1$ \citep{Hebb2009}, with an
uncertainty multiplier of 2 (used to inflate the stellar limb darkening profile
uncertainty estimates).

The blending analysis results are shown in Fig.~\ref{fig:wasp12b-posteriors}. The marginal 
contamination posterior mode agrees with the true reference passband contamination value of 
$4\%$, but contamination levels up to $\sim 25\%$ are allowed. However, looking at the
joint distributions, it is clear that significant contamination from sources with $\teffh \ne \teffc$
is excluded. Interestingly, also the low-contamination solutions by a contaminant without a colour 
difference are ruled out, and a correct lower-temperature contaminant is favoured instead.

\section{Discussion and Conclusions}
\label{sec:conclusions}
Multicolour transit photometry combined with a physics-based contamination model
can be used to estimate the true radius ratio of the transiting object in the presence 
of possible flux contamination from unresolved sources, even from the transiting object
itself. The true radius ratio combined with a stellar radius estimate results in an absolute
radius estimate that can be used to validate a transiting object as a planet if its radius
is smaller than 0.8~\rjup, the lower limit of what is expected for brown dwarfs 
\citep{Burrows2011,Chabrier2000}. 

The visualisation of the results from the analyses (simulations and observations) 
presented in this paper focused on showing the estimated contamination level in an arbitrarily 
chosen reference passband. However, the true radius ratio, and especially its upper limit, is the
main quantity of interest in planet candidate validation. 

The apparent radius 
ratio is generally well-constrained by the observations ($\para \sim \sqrt{\Delta F}$), 
but the true radius ratio posterior depends also on how well the photometry can constrain
the contamination space. As detailed in Appendix~\ref{sec:true_rr_and_contamination},
(and visible in the simulations) decreasing signal to noise (increasing uncertainty in \para) 
leads to a contamination posterior with an increasingly non-zero mode. However, the contamination
posterior should always be considered in the context of the whole parameter space (as illustrated
in the joint-parameter plots), and the maximum contamination allowed by the observations
(giving the maximum \ptra) should be considered as the main output of the analysis.

The required photometric precision for multicolour transit
photometry to be used efficiently in planet candidate validation can be reached with small
($\sim 1$~m) ground-based telescopes. Further, the current developments in ground-based
photometry, such as the use of diffusers \citep{Stefansson2017}, can improve the achievable
precision to level where only a few transits are required for candidate validation.

In practice, the analysis can be made more robust by comparing
the results with analyses of simulations mimicking the observations. This is
feasible since a single analysis of multiple transits observed in multiple colours (real 
observations or simulations) takes minutes, and can give insight into the possible biases
arising from the signal to noise ratio or observing geometry.

In some cases, this technique can confirm candidates with less telescope time than the RV
technique, particularly for long period systems. While the mass
is certainly a desirable parameter to possess, when in a situation with a large backlog of
candidates requiring verification, the less time needed on "premium" telescope facilities
for confirmation, the better. Moreover, the intrinsic RV noise in stars, orbital
eccentricity, and the presence of unknown planets can complicate attempts to confirm
candidates with RV. From space, it should be possible to confirm low-mass, long period
planets that would be enormously difficult to impossible to confirm with RV.

The method does have limitations: it cannot separate grazing exoplanets from grazing
binary stars (unless blended, in which case different depths in different colours are
still likely), nor can it distinguish between red dwarfs, brown dwarfs and giant planets
-- in these cases, an upper mass limit from RV is necessary. However, this is not
necessarily a severe problem. Transiting exoplanet surveys have found that the red and brown
dwarfs seem to be much more rare than transiting exoplanets despite the fact that they
induced much stronger RV variations -- e.g. out of the 22 announced \corot exoplanets, only
two (CoRoT-3b, CoRoT-15b) seem to be brown dwarfs. On the other hand, \citet{Santerne2012}
found that approximately 35\% of close-in giant planet candidates are non-planetary false
positives. Half of these were found to be CEBs after extensive RV follow-up and likely
would have been more easily identified as such with multicolour photometry.

Further, transits with high impact parameters (not necessarily grazing) lead to
overestimated contamination. This is connected to the fact that increasing uncertainty in 
the true radius ratio leads to a bias in the contamination, as detailed in Appendix~\ref{sec:true_rr_and_contamination}.
For a high  impact parameter, even the apparent radius ratio can be poorly constrained, 
and the true radius ratio even more so, which leads to systematically high contamination 
estimates.

The approach is very good at identifying
CEBs \citep{ODonovan2006,ODonovan2007}, which are
the most problematic cases for the RV technique; indeed, many hours of
very valuable telescope time have been wasted on fruitless efforts to
detect the RV signal of a possible planet which was actually a CEB.  In
some cases, multicolour photometry may actually be the best technique
for follow-up -- not only for relatively deep transits ($\gtrsim
0.8\%$) with long periods or faint $V \gtrsim15$ host stars using ground-based
photometry but also for earth-size candidates using space telescopes.
One practical application of this technique would be a systematic effort
to follow up \tess candidates. Combining these observations 
with the \tess light curves would allow relatively efficient characterisation 
of such systems. 

Naturally, the method is applicable only when the
candidate signal is due to a real astrophysical phenomenon, such as transiting planet
or eclipsing binary, and not due to an instrumental artefact (also a source of false
candidates). However, these candidates can be directly rejected based on the lack of
a transit in the follow-up photometry.

The strength of
the colour difference depends not only on limb darkening and transit
parameters, but also on the colour difference between the light
sources, arising from temperature differences and interstellar reddening.
Despite this, it would be eminently practical for future satellite missions
dedicated to have the capability to observe in multiple colours. At the cost
of some photons, a large fraction of the telescope resources committed to
the follow-up effort would be spared, as RV follow up would only be
necessary in case when the planetary mass was of particular scientific
interest.

\begin{acknowledgements}
  We thank the anonymous referee for their helpful comments.
  HD acknowledges support by grants ESP2015-65712-C5-4-R and ESP2017-87676-C5-4-R,
  and FM, RA, and EP by grant ESP2016-80435-C2-2-R, all from the Spanish Ministry of
  Economy and Competitiveness (MINECO/MICINN). VJSB acknowledges support by grant 
  AYA2015- 69350-C3-2-P. NN acknowledges supports by JSPS KAKENHI Grant Numbers JP17H04574, 
  JP18H01265 and JP18H05439, and JST PRESTO Grant Number JPMJPR1775. This article is 
  partly based on observations made with the MuSCAT2 instrument, developed by ABC, 
  at Telescopio Carlos S\'anchez operated on the island of Tenerife by the IAC in the 
  Spanish Observatorio del Teide. The authors wish to acknowledge the contribution of Teide 
  High-Performance Computing facilities to the results of this research.
  TeideHPC facilities are provided by the Instituto Tecnológico 
  y de Energías Renovables (ITER, SA). URL: http://teidehpc.iter.es.  The authors thankfully 
  acknowledges the technical expertise and assistance provided by the 
  Spanish Supercomputing Network (Red Española de Supercomputación), as well as the computer 
  resources used: the LaPalma Supercomputer, located at the Instituto de Astrofísica de Canarias.
  \end{acknowledgements}

\bibliographystyle{aa} 
\bibliography{colsig}

\begin{thebibliography}{66}
\expandafter\ifx\csname natexlab\endcsname\relax\def\natexlab#1{#1}\fi

\bibitem[{Almenara {et~al.}(2009)Almenara, Deeg, Aigrain, Alonso, Auvergne,
  Baglin, Barbieri, Barge, Bord{\'{e}}, Bouchy, Bruntt, Cabrera, Carone,
  Carpano, Catala, Csizmadia, {De la Reza}, Deleuil, Dvorak, Erikson, Fridlund,
  Gandolfi, Gillon, Gondoin, Guenther, Guillot, Hatzes, H{\'{e}}brard, Jorda,
  Lammer, L{\'{e}}ger, Llebaria, Loeillet, Magain, Mayor, Mazeh, Moutou,
  Ollivier, P{\"{a}}tzold, Pont, Queloz, Rauer, R{\'{e}}gulo, Renner, Rouan,
  Samuel, Schneider, Shporer, Wuchterl, \& Zucker}]{Almenara2009}
Almenara, J.~M., Deeg, H.~J., Aigrain, S., {et~al.} 2009, A{\&}A, 506, 337

\bibitem[{Ambikasaran {et~al.}(2014)Ambikasaran, Foreman-Mackey, Greengard,
  Hogg, \& O'Neil}]{Ambikasaran2014}
Ambikasaran, S., Foreman-Mackey, D., Greengard, L., Hogg, D.~W., \& O'Neil, M.
  2014 [\eprint[arXiv]{1403.6015}]

\bibitem[{Armstrong {et~al.}(2017)Armstrong, Pollacco, \&
  Santerne}]{Armstrong2017}
Armstrong, D.~J., Pollacco, D., \& Santerne, A. 2017, MNRAS, 465, 2634

\bibitem[{Ballard {et~al.}(2011)Ballard, Fabrycky, Fressin, Charbonneau,
  Desert, Torres, Marcy, Burke, Isaacson, Henze, Steffen, Ciardi, Howell,
  Cochran, Endl, Bryson, Rowe, Holman, Lissauer, Jenkins, Still, Ford,
  Christiansen, Middour, Haas, Li, Hall, McCauliff, Batalha, Koch, \&
  Borucki}]{Ballard2011a}
Ballard, S., Fabrycky, D., Fressin, F., {et~al.} 2011, ApJ, 743, 200

\bibitem[{Batalha {et~al.}(2010)Batalha, Rowe, Gilliland, Jenkins, Caldwell,
  Borucki, Koch, Lissauer, Dunham, Gautier, Howell, Latham, Marcy, \&
  Prsa}]{Batalha2010}
Batalha, N.~M., Rowe, J.~F., Gilliland, R.~L., {et~al.} 2010, ApJ, 713, L103

\bibitem[{Bechter {et~al.}(2014)Bechter, Crepp, Ngo, Knutson, Batygin, Hinkley,
  Muirhead, Johnson, Howard, Montet, Matthews, \& Morton}]{Bechter2013}
Bechter, E.~B., Crepp, J.~R., Ngo, H., {et~al.} 2014, Astrophys. J., 788, 2

\bibitem[{Bradley {et~al.}(2019)Bradley, Sipocz, Robitaille, Tollerud,
  Vin{\'{i}}cius, Deil, Barbary, G{\"{u}}nther, Cara, Busko, Conseil,
  Droettboom, Bostroem, Bray, Bratholm, Wilson, Craig, Barentsen, Pascual,
  Donath, Greco, Perren, Lim, \& Kerzendorf}]{Bradley2019}
Bradley, L., Sipocz, B., Robitaille, T., {et~al.} 2019

\bibitem[{Brown(2003)}]{Brown2003}
Brown, T.~M. 2003, ApJ, 593, L125

\bibitem[{Bryson {et~al.}(2013)Bryson, Jenkins, Gilliland, Twicken, Clarke,
  Rowe, Caldwell, Batalha, Mullally, Haas, \& Tenenbaum}]{Bryson2013}
Bryson, S.~T., Jenkins, J.~M., Gilliland, R.~L., {et~al.} 2013, Publ. Astron.
  Soc. Pacific, 125, 889

\bibitem[{Burrows {et~al.}(2011)Burrows, Heng, \& Nampaisarn}]{Burrows2011}
Burrows, A.~S., Heng, K., \& Nampaisarn, T. 2011, Astrophys. J., 736, 47

\bibitem[{Cabrera {et~al.}(2017)Cabrera, Barros, Armstrong, Hidalgo, Santos,
  Almenara, Alonso, Deleuil, Demangeon, D{\'{i}}az, Lendl, Pfaff, Rauer,
  Santerne, Serrano, \& Zucker}]{Cabrera2017a}
Cabrera, J., Barros, S. C.~C., Armstrong, D., {et~al.} 2017, A{\&}A, 606, A75

\bibitem[{Cameron(2012)}]{Cameron2012}
Cameron, A.~C. 2012, Nature, 492, 48

\bibitem[{Carone {et~al.}(2012)Carone, Gandolfi, Cabrera, Hatzes, Deeg,
  Csizmadia, P{\"{a}}tzold, Weingrill, Aigrain, Alonso, Alapini, Almenara,
  Auvergne, Baglin, Barge, Bonomo, Bord{\'{e}}, Bouchy, Bruntt, Carpano,
  Cochran, Deleuil, D{\'{i}}az, Dreizler, Dvorak, Eisl{\"{o}}ffel,
  Eigm{\"{u}}ller, Endl, Erikson, Ferraz-Mello, Fridlund, Gazzano, Gibson,
  Gillon, Gondoin, Grziwa, G{\"{u}}nther, Guillot, Hartmann, Havel,
  H{\'{e}}brard, Jorda, Kabath, L{\'{e}}ger, Llebaria, Lammer, Lovis, MacQueen,
  Mayor, Mazeh, Moutou, Nortmann, Ofir, Ollivier, Parviainen, Pepe, Pont,
  Queloz, Rabus, Rauer, R{\'{e}}gulo, Renner, de~la Reza, Rouan, Santerne,
  Samuel, Schneider, Shporer, Stecklum, Tal-Or, Tingley, Udry, \&
  Wuchterl}]{Carone2011}
Carone, L., Gandolfi, D., Cabrera, J., {et~al.} 2012, A{\&}A, 538, A112

\bibitem[{Chabrier \& Baraffe(2000)}]{Chabrier2000}
Chabrier, G. \& Baraffe, I. 2000, Annu. Rev. Astron. Astrophys., 38, 337

\bibitem[{Chen {et~al.}(2014)Chen, van Boekel, Wang, Nikolov, Fortney, Seemann,
  Wang, Mancini, \& Henning}]{Chen2014}
Chen, G., van Boekel, R., Wang, H., {et~al.} 2014, A{\&}A, 563, A40

\bibitem[{Cochran {et~al.}(2011)Cochran, Fabrycky, Torres, Fressin,
  D{\'{e}}sert, Ragozzine, Sasselov, Fortney, Rowe, Brugamyer, Bryson, Carter,
  Ciardi, Howell, Steffen, Borucki, Koch, Winn, Welsh, Uddin, Tenenbaum, Still,
  Seager, Quinn, Mullally, Miller, Marcy, MacQueen, Lucas, Lissauer, Latham,
  Knutson, Kinemuchi, Johnson, Jenkins, Isaacson, Howard, Horch, Holman, Henze,
  Haas, Gilliland, {Gautier III}, Ford, Fischer, Everett, Endl, Demory, Deming,
  Charbonneau, Caldwell, Buchhave, Brown, \& Batalha}]{Cochran2011a}
Cochran, W.~D., Fabrycky, D.~C., Torres, G., {et~al.} 2011, Astrophys. J.
  Suppl. Ser., 197, 7

\bibitem[{Col{\'{o}}n \& Ford(2011)}]{Colon2011}
Col{\'{o}}n, K.~D. \& Ford, E.~B. 2011, PASP, 123, 1391

\bibitem[{Coughlin {et~al.}(2014)Coughlin, Thompson, Bryson, Burke, Caldwell,
  Christiansen, Haas, Howell, Jenkins, Kolodziejczak, Mullally, \&
  Rowe}]{Coughlin2014}
Coughlin, J.~L., Thompson, S.~E., Bryson, S.~T., {et~al.} 2014, Astron. J.,
  147, 119

\bibitem[{Crossfield {et~al.}(2012)Crossfield, Hansen, \&
  Barman}]{Crossfield2012}
Crossfield, I. J.~M., Hansen, B. M.~S., \& Barman, T. 2012, Astrophys. J., 746,
  46

\bibitem[{Daemgen {et~al.}(2009)Daemgen, Hormuth, Brandner, Bergfors, Janson,
  Hippler, \& Henning}]{Daemgen2009}
Daemgen, S., Hormuth, F., Brandner, W., {et~al.} 2009, A{\&}A, 498, 567

\bibitem[{Deeg {et~al.}(2009)Deeg, Gillon, Shporer, Rouan, Stecklum, Aigrain,
  Alapini, Almenara, Alonso, Barbieri, Bouchy, Eisl{\"{o}}ffel, Erikson,
  Fridlund, Eigm{\"{u}}ller, Handler, Hatzes, Kabath, Lendl, Mazeh, Moutou,
  Queloz, Rauer, Rabus, Tingley, \& Titz}]{Deeg2009}
Deeg, H.~J., Gillon, M., Shporer, A., {et~al.} 2009, A{\&}A, 506, 343

\bibitem[{Deeg \& Tingley(2017)}]{Deeg2016}
Deeg, H.~J. \& Tingley, B. 2017, A{\&}A, 599, A93

\bibitem[{D{\'{i}}az {et~al.}(2014)D{\'{i}}az, Almenara, Santerne, Moutou,
  Lethuillier, \& Deleuil}]{Diaz2014}
D{\'{i}}az, R.~F., Almenara, J.~M., Santerne, A., {et~al.} 2014, MNRAS, 441,
  983

\bibitem[{Drake(2003)}]{Drake2003}
Drake, A.~J. 2003, Astrophys. J., 589, 1020

\bibitem[{Foreman-Mackey {et~al.}(2013)Foreman-Mackey, Hogg, Lang, \&
  Goodman}]{Foreman-Mackey2012}
Foreman-Mackey, D., Hogg, D.~W., Lang, D., \& Goodman, J. 2013, Publ. Astron.
  Soc. Pacific, 125, 306

\bibitem[{Fressin {et~al.}(2013)Fressin, Torres, Charbonneau, Bryson,
  Christiansen, Dressing, Jenkins, Walkowicz, \& Batalha}]{Fressin2013}
Fressin, F., Torres, G., Charbonneau, D., {et~al.} 2013, Astrophys. J., 766, 81

\bibitem[{Gillon {et~al.}(2012)Gillon, Triaud, Fortney, Demory, Jehin, Lendl,
  Magain, Kabath, Queloz, Alonso, Anderson, {Collier Cameron}, Fumel, Hebb,
  Hellier, Lanotte, Maxted, Mowlavi, \& Smalley}]{Gillon2012}
Gillon, M., Triaud, A. H. M.~J., Fortney, J.~J., {et~al.} 2012, A{\&}A, 542, A4

\bibitem[{Gim{\'{e}}nez(2006)}]{Gimenez2006}
Gim{\'{e}}nez, A. 2006, Astrophys. J., 650, 408

\bibitem[{Goodman \& Weare(2010)}]{Goodman2010}
Goodman, J. \& Weare, J. 2010, Commun. Appl. Math. Comput. Sci., 5, 65

\bibitem[{Guenther {et~al.}(2013)Guenther, Fridlund, Alonso, Carpano, Deeg,
  Deleuil, Dreizler, Endl, Gandolfi, Gillon, Guillot, Jehin, L{\'{e}}ger,
  Moutou, Nortmann, Rouan, Samuel, Schneider, \& Tingley}]{Guenther2013}
Guenther, E.~W., Fridlund, M., Alonso, R., {et~al.} 2013, A{\&}A, 556, A75

\bibitem[{Hebb {et~al.}(2009)Hebb, Collier-Cameron, Loeillet, Pollacco,
  H{\'{e}}brard, Street, Bouchy, Stempels, Moutou, Simpson, Udry, Joshi, West,
  Skillen, Wilson, McDonald, Gibson, Aigrain, Anderson, Benn, Christian, Enoch,
  Haswell, Hellier, Horne, Irwin, Lister, Maxted, Mayor, Norton, Parley, Pont,
  Queloz, Smalley, \& Wheatley}]{Hebb2009}
Hebb, L., Collier-Cameron, A., Loeillet, B., {et~al.} 2009, Astrophys. J., 693,
  1920

\bibitem[{Hellier {et~al.}(2011)Hellier, Anderson, Cameron, Gillon, Jehin,
  Lendl, Maxted, Pepe, Pollacco, Queloz, Segransan, Smalley, Smith, Southworth,
  Triaud, Udry, \& West}]{Hellier2011}
Hellier, C., Anderson, D.~R., Cameron, A.~C., {et~al.} 2011, A{\&}A, 4

\bibitem[{Hunter(2007)}]{Hunter2007}
Hunter, J.~D. 2007, Comput. Sci. Eng., 9, 90

\bibitem[{Husser {et~al.}(2013)Husser, {Wende-von Berg}, Dreizler, Homeier,
  Reiners, Barman, \& Hauschildt}]{Husser2013}
Husser, T.-O., {Wende-von Berg}, S., Dreizler, S., {et~al.} 2013, A{\&}A, 553,
  A6

\bibitem[{Lang {et~al.}(2010)Lang, Hogg, Mierle, Blanton, \& Roweis}]{Lang2010}
Lang, D., Hogg, D.~W., Mierle, K., Blanton, M., \& Roweis, S. 2010, Astron. J.,
  139, 1782

\bibitem[{Mandushev {et~al.}(2005)Mandushev, Torres, Latham, Charbonneau,
  Alonso, White, Stefanik, Dunham, Brown, \& O'Donovan}]{Mandushev2005}
Mandushev, G., Torres, G., Latham, D.~W., {et~al.} 2005, ApJ, 621, 1061

\bibitem[{Mckinney(2010)}]{Mckinney2010}
Mckinney, W. 2010in , 51--56

\bibitem[{Morton {et~al.}(2016)Morton, Bryson, Coughlin, Rowe, Ravichandran,
  Petigura, Haas, \& Batalha}]{Morton2016}
Morton, T.~D., Bryson, S.~T., Coughlin, J.~L., {et~al.} 2016, Astrophys. J.,
  822, 86

\bibitem[{Moutou {et~al.}(2009)Moutou, Pont, Bouchy, Deleuil, Almenara, Alonso,
  Barbieri, Bruntt, Deeg, Fridlund, Gandolfi, Gillon, Guenther, Hatzes,
  H{\'{e}}brard, Loeillet, Mayor, Mazeh, Queloz, Rabus, Rouan, Shporer, Udry,
  Aigrain, Auvergne, Baglin, Barge, Benz, Bord{\'{e}}, Carpano, {De la Reza},
  Dvorak, Erikson, Gondoin, Guillot, Jorda, Kabath, Lammer, L{\'{e}}ger,
  Llebaria, Lovis, Magain, Ollivier, P{\"{a}}tzold, Pepe, Rauer, Schneider, \&
  Wuchterl}]{Moutou2009}
Moutou, C., Pont, F., Bouchy, F., {et~al.} 2009, A{\&}A, 506, 321

\bibitem[{Mullally {et~al.}(2016)Mullally, Coughlin, Thompson, Christiansen,
  Burke, Clarke, \& Haas}]{Mullally2016}
Mullally, F., Coughlin, J.~L., Thompson, S.~E., {et~al.} 2016, Publ. Astron.
  Soc. Pacific, 128, 074502

\bibitem[{Mullally {et~al.}(2018)Mullally, Thompson, Coughlin, Burke, \&
  Rowe}]{Mullally2018}
Mullally, F., Thompson, S.~E., Coughlin, J.~L., Burke, C.~J., \& Rowe, J.~F.
  2018, Astron. J., 155, 210

\bibitem[{Murgas {et~al.}(2014)Murgas, Pall{\'{e}}, {Zapatero Osorio},
  Nortmann, Hoyer, \& Cabrera-Lavers}]{Murgas2014}
Murgas, F., Pall{\'{e}}, E., {Zapatero Osorio}, M.~R., {et~al.} 2014, A{\&}A,
  563, A41

\bibitem[{Narita {et~al.}(2018)Narita, Fukui, Kusakabe, Watanabe, Palle,
  Parviainen, Monta{\~{n}}{\'{e}}s-Rodr{\'{i}}guez, Murgas, Monelli, Aguiar, \&
  {Perez Prieto}}]{Narita2018}
Narita, N., Fukui, A., Kusakabe, N., {et~al.} 2018, J. Astron. Telesc.
  Instruments, Syst., 5, 1

\bibitem[{O'Donovan {et~al.}(2007)O'Donovan, Charbonneau, Alonso, Brown,
  Mandushev, Dunham, Latham, Stefanik, Torres, \& Everett}]{ODonovan2007}
O'Donovan, F.~T., Charbonneau, D., Alonso, R., {et~al.} 2007, Astrophys. J.,
  662, 658

\bibitem[{O'Donovan {et~al.}(2006)O'Donovan, Charbonneau, Torres, Mandushev,
  Dunham, Latham, Alonso, Brown, Esquerdo, Everett, \& Creevey}]{ODonovan2006}
O'Donovan, F.~T., Charbonneau, D., Torres, G., {et~al.} 2006, Astrophys. J.,
  644, 1237

\bibitem[{Parviainen(2015)}]{Parviainen2015}
Parviainen, H. 2015, MNRAS, 450, 3233

\bibitem[{Parviainen(2018)}]{Parviainen2018}
Parviainen, H. 2018, in Handb. Exopl. (Cham: Springer International
  Publishing), 1--24

\bibitem[{Parviainen \& Aigrain(2015)}]{Parviainen2015b}
Parviainen, H. \& Aigrain, S. 2015, MNRAS, 453, 3822

\bibitem[{Perez \& Granger(2007)}]{Perez2007}
Perez, F. \& Granger, B. 2007, Comput. Sci. Eng., 21

\bibitem[{Peterson(2009)}]{Peterson2009}
Peterson, P. 2009, Int. J. Comput. Sci. Eng., 4, 296

\bibitem[{Pollacco {et~al.}(2006)Pollacco, Skillen, Cameron, Christian,
  Hellier, Irwin, Lister, Street, West, Anderson, Clarkson, Deeg, Enoch, Evans,
  Fitzsimmons, Haswell, Hodgkin, Horne, Kane, Keenan, Maxted, Norton, Osborne,
  Parley, Ryans, Smalley, Wheatley, \& Wilson}]{Pollacco2006}
Pollacco, D.~L., Skillen, I., Cameron, A.~C., {et~al.} 2006, PASP, 118, 1407

\bibitem[{Quintana {et~al.}(2013)Quintana, Rowe, Barclay, Howell, Ciardi,
  Demory, Caldwell, Borucki, Christiansen, Jenkins, Klaus, Fulton, Morris,
  Sanderfer, Shporer, Smith, Still, \& Thompson}]{Quintana2013}
Quintana, E.~V., Rowe, J.~F., Barclay, T.~S., {et~al.} 2013, Astrophys. J.,
  767, 137

\bibitem[{Rauer {et~al.}(2014)Rauer, Catala, Aerts, Appourchaux, Benz,
  Brandeker, Christensen-Dalsgaard, Deleuil, Gizon, Goupil, G{\"{u}}del,
  Janot-Pacheco, Mas-Hesse, Pagano, Piotto, Pollacco, Santos, Smith,
  Su{\'{a}}rez, Szab{\'{o}}, Udry, Adibekyan, Alibert, Almenara, Amaro-Seoane,
  Eiff, Asplund, Antonello, Barnes, Baudin, Belkacem, Bergemann, Bihain, Birch,
  Bonfils, Boisse, Bonomo, Borsa, Brand{\~{a}}o, Brocato, Brun, Burleigh,
  Burston, Cabrera, Cassisi, Chaplin, Charpinet, Chiappini, Church, Csizmadia,
  Cunha, Damasso, Davies, Deeg, D{\'{i}}az, Dreizler, Dreyer, Eggenberger,
  Ehrenreich, Eigm{\"{u}}ller, Erikson, Farmer, Feltzing, de~{Oliveira Fialho},
  Figueira, Forveille, Fridlund, Garc{\'{i}}a, Giommi, Giuffrida, Godolt,
  da~Silva, Granzer, Grenfell, Grotsch-Noels, G{\"{u}}nther, Haswell, Hatzes,
  H{\'{e}}brard, Hekker, Helled, Heng, Jenkins, Johansen, Khodachenko,
  Kislyakova, Kley, Kolb, Krivova, Kupka, Lammer, Lanza, Lebreton, Magrin,
  Marcos-Arenal, Marrese, Marques, Martins, Mathis, Mathur, Messina, Miglio,
  Montalban, Montalto, {P. F. G. Monteiro}, Moradi, Moravveji, Mordasini,
  Morel, Mortier, Nascimbeni, Nelson, Nielsen, Noack, Norton, Ofir, Oshagh,
  Ouazzani, P{\'{a}}pics, Parro, Petit, Plez, Poretti, Quirrenbach, Ragazzoni,
  Raimondo, Rainer, Reese, Redmer, Reffert, Rojas-Ayala, Roxburgh, Salmon,
  Santerne, Schneider, Schou, Schuh, Schunker, Silva-Valio, Silvotti, Skillen,
  Snellen, Sohl, Sousa, Sozzetti, Stello, Strassmeier, {\v{S}}vanda,
  Szab{\'{o}}, Tkachenko, Valencia, {Van Grootel}, Vauclair, Ventura, Wagner,
  Walton, Weingrill, Werner, Wheatley, \& Zwintz}]{Rauer2013}
Rauer, H., Catala, C., Aerts, C., {et~al.} 2014, Exp. Astron., 38, 249

\bibitem[{Ricker {et~al.}(2014)Ricker, Winn, Vanderspek, Latham, Bakos, Bean,
  Berta-Thompson, Brown, Buchhave, Butler, Butler, Chaplin, Charbonneau,
  Christensen-Dalsgaard, Clampin, Deming, Doty, {De Lee}, Dressing, Dunham,
  Endl, Fressin, Ge, Henning, Holman, Howard, Ida, Jenkins, Jernigan, Johnson,
  Kaltenegger, Kawai, Kjeldsen, Laughlin, Levine, Lin, Lissauer, MacQueen,
  Marcy, McCullough, Morton, Narita, Paegert, Palle, Pepe, Pepper, Quirrenbach,
  Rinehart, Sasselov, Sato, Seager, Sozzetti, Stassun, Sullivan, Szentgyorgyi,
  Torres, Udry, \& Villasenor}]{Ricker2014}
Ricker, G.~R., Winn, J.~N., Vanderspek, R., {et~al.} 2014, 914320

\bibitem[{Rosenblatt(1971)}]{Rosenblatt1971}
Rosenblatt, F. 1971, Icarus, 14, 71

\bibitem[{Santerne {et~al.}(2015)Santerne, D{\'{i}}az, Almenara, Bouchy,
  Deleuil, Figueira, H{\'{e}}brard, Moutou, Rodionov, \& Santos}]{Santerne2015}
Santerne, A., D{\'{i}}az, R.~F., Almenara, J.-M., {et~al.} 2015, MNRAS, 451,
  2337

\bibitem[{Santerne {et~al.}(2012)Santerne, D{\'{i}}az, Moutou, Bouchy,
  H{\'{e}}brard, Almenara, Bonomo, Deleuil, \& Santos}]{Santerne2012}
Santerne, A., D{\'{i}}az, R.~F., Moutou, C., {et~al.} 2012, A{\&}A, 545, A76

\bibitem[{Stefansson {et~al.}(2017)Stefansson, Mahadevan, Hebb, Wisniewski,
  Huehnerhoff, Morris, Halverson, Zhao, Wright, O'rourke, Knutson, Hawley,
  Kanodia, Li, Hagen, Liu, Beatty, Bender, Robertson, Dembicky, Gray,
  Ketzeback, McMillan, \& Rudyk}]{Stefansson2017}
Stefansson, G., Mahadevan, S., Hebb, L., {et~al.} 2017, Astrophys. J., 848, 9

\bibitem[{{The Astropy Collaboration} {et~al.}(2013){The Astropy
  Collaboration}, Robitaille, Tollerud, Greenfield, Droettboom, Bray, Aldcroft,
  Davis, Ginsburg, Price-Whelan, Kerzendorf, Conley, Crighton, Barbary, Muna,
  Ferguson, Grollier, Parikh, Nair, G{\"{u}}nther, Deil, Woillez, Conseil,
  Kramer, Turner, Singer, Fox, Weaver, Zabalza, Edwards, Bostroem, Burke,
  Casey, Crawford, Dencheva, Ely, Jenness, Labrie, Lim, Pierfederici, Pontzen,
  Ptak, Refsdal, Servillat, \& Streicher}]{TheAstropyCollaboration2013}
{The Astropy Collaboration}, Robitaille, T.~P., Tollerud, E.~J., {et~al.} 2013,
  33, 1

\bibitem[{Tingley(2004)}]{Tingley2004}
Tingley, B. 2004, A{\&}A, 425, 1125

\bibitem[{Tingley {et~al.}(2011)Tingley, Palle, Parviainen, Deeg, Osorio,
  Cabrera-Lavers, Belmonte, Rodriguez, Murgas, \& Ribas}]{Tingley2011b}
Tingley, B., Palle, E., Parviainen, H., {et~al.} 2011, Astron. Astrophys.
  Lett., 536, 9

\bibitem[{Tingley {et~al.}(2014)Tingley, Parviainen, Gandolfi, Deeg, Palle,
  {Monta{\~{n}}{\'{e}}s Rodriguez}, Murgas, Alonso, Bruntt, \&
  Fridlund}]{Tingley2014a}
Tingley, B., Parviainen, H., Gandolfi, D., {et~al.} 2014, A{\&}A, 567
  [\eprint[arXiv]{arXiv:1405.5354v2}]

\bibitem[{Torres {et~al.}(2011)Torres, Fressin, Batalha, Borucki, Brown,
  Bryson, Buchhave, Charbonneau, Ciardi, Dunham, Fabrycky, Ford, {Gautier III},
  Gilliland, Holman, Howell, Isaacson, Jenkins, Koch, Latham, Lissauer, Marcy,
  Monet, Prsa, Quinn, Ragozzine, Rowe, Sasselov, Steffen, \&
  Welsh}]{Torres2011}
Torres, G., Fressin, F., Batalha, N.~M., {et~al.} 2011, ApJ, 727, 24

\bibitem[{van~der Walt {et~al.}(2011)van~der Walt, Colbert, \&
  Varoquaux}]{VanderWalt2011}
van~der Walt, S., Colbert, S.~C., \& Varoquaux, G. 2011, Comput. Sci. Eng., 13,
  22

\bibitem[{von Essen {et~al.}(2019)von Essen, Stefansson, Mallonn, Pursimo,
  Djupvik, Mahadevan, Kjeldsen, Freudenthal, \& Dreizler}]{VonEssen2019}
von Essen, C., Stefansson, G., Mallonn, M., {et~al.} 2019
  [\eprint[arXiv]{1904.05362}]

\bibitem[{Worek(2000)}]{Worek2000}
Worek, T. 2000, Inf. Bull. Var. Stars, 1

\end{thebibliography}

\appendix
\section{Contaminated transit model}
\label{sec:appendix.model}

The light contamination model in \pytransit~v2 depicts the observed flux as a
combination of the host star flux and the contaminant flux (possibly from multiple
contaminating sources). The contamination is calculated for a set of passbands 
given the passband transmission functions, host and contaminant effective temperatures, 
and the level of contamination in some reference passband.

\pytransit offers two contamination models: a simple black-body model where the stars are
approximated as black bodies, and a more realistic model using PHOENIX-calculated stellar
spectra by \citet{Husser2013}. In the latter, the spectra calculated for a ($\teff,\, Z,\,
\log\;g$) grid  are  downsampled to a lower spectral resolution, and averaged over $Z$ and
$\log\;g$. The set of averaged downsampled spectra is stored as a single 2D array, from
which the stellar fluxes can be interpolated for $2300~\mathrm{K} < \teff <
12\,000~\mathrm{K}$ and $300~\mathrm{nm} < \lambda < 1\,000~\mathrm{nm}$.

The observed flux is a linear combination of host and contaminant star fluxes, possibly
from several contaminating stars. The contaminated model flux, $F_c$, for a single
passband normalised to the out-of-transit flux level is
\begin{equation}
	F_c = c + (1-c) F_0, \qquad 0 \leq c \leq 1,
\end{equation}
where $c$ is the contamination fraction, and $F_0$ the (uncontaminated) host-star flux.
The transit depth scales linearly with the contamination factor, $\Delta F_c = \Delta F_0
(1-c)$, so the apparent and true planet-star radius ratios are 
\begin{equation}
    \para = \ptra \sqrt{1-c}, \qquad  \ptra = \para / \sqrt{1-c},
\end{equation}
respectively. 

\begin{figure}
	\centering
	\includegraphics[width=\columnwidth]{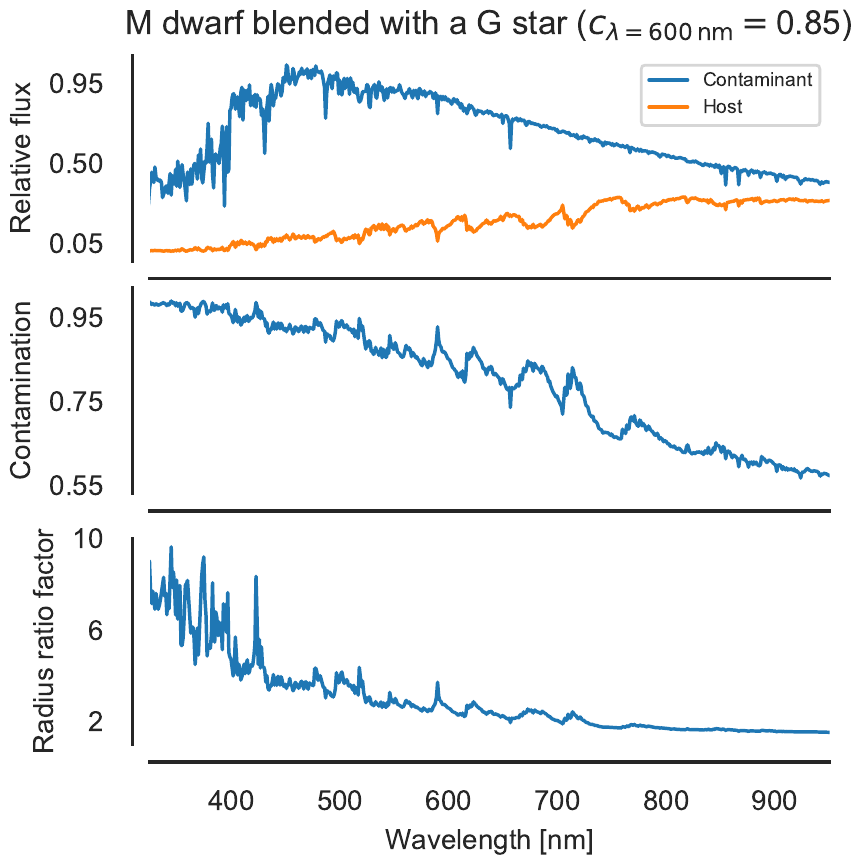}
	\caption{A $\teff=3600$~K M dwarf strongly contaminated ($c_{\lambda=600\,\text{nm}}=0.85$) by a $\teff=5800$~K G star. The upper panel shows the relative stellar fluxes as a function of wavelength,
	the middle panel shows the contamination factor as a function of wavelength, and the bottom panel shows the radius ratio factor ($\ptra/\para$) as a function
	of wavelength.}
	\label{fig:mgcontamination}
\end{figure}

The relative host and contaminant fluxes ($F_\mathrm{H}$ and $F_\mathrm{C}$, respectively)
for wavelength $\lambda$ are
\begin{equation}
	F_\mathrm{H}(\lambda) = (1-c_0) \frac{S(\teffh, \lambda)}{S(\teffh, \lambda_0)}, \quad F_\mathrm{C}(\lambda) = c_0 \frac{S(\teffc, \lambda)}{S(\teffc, \lambda_0)},
\end{equation} 
where $S$ is the flux model, $\lambda_0$ is a reference wavelength, $c_0$ is the contamination level in this
reference wavelength, and \teffh and \teffc are the host and contaminant effective
temperatures, respectively. Now, contamination for any given wavelength is
\begin{equation}
	c(\lambda) = \frac{F_\mathrm{C}(\lambda)}{F_\mathrm{H}(\lambda) + F_\mathrm{C}(\lambda)},
\end{equation}
as illustrated in Fig.~\ref{fig:mgcontamination}.

The relative fluxes integrated over a set of passbands defined by transmission functions
$\mathcal{T}$,  are obtained by multiplying the flux model by the transmission function and
integrating over the wavelength, as
\begin{equation}
F_\mathrm{H,i} = (1-c_0) \frac{\int \mathcal{T}_i (\lambda)\;  S(\teffh, \lambda)\; \ud\lambda}{\int \mathcal{T}_0(\lambda)\; S(\teffh, \lambda)\; \ud\lambda}, 
\end{equation} 
\begin{equation}
F_\mathrm{C,i} = c_0 \frac{\int \mathcal{T}_i (\lambda)\;  S(\teffc, \lambda)\; \ud\lambda}{\int \mathcal{T}_0(\lambda)\; S(\teffc, \lambda)\; \ud\lambda}, 
\end{equation} 
where $c_0$ is now the amount of contamination in the reference passband.
The contamination for passband $i$ is now
\begin{equation}
c_i = \frac{F_\mathrm{C,i}}{F_\mathrm{H,i} + F_\mathrm{C,i}}.
\end{equation}.

The main science case for the  contamination model is in multicolour transit candidate
analysis, as detailed earlier in this paper. Combining the contamination model with a
transit model allows one to estimate the true, uncontaminated, radius ratio of a
transiting exoplanet. The model has been integrated into \pytransit~v2, and an
\textsc{IPython} notebook  tutorial on how to use the model in practical analysis can be
found from \url{github.com/hpparvi/PyTransit/notebooks}.

\section{True radius ratio and contamination}
\label{sec:true_rr_and_contamination}

The flux contamination can be derived from the apparent and true radius ratios as $1 - \paaa/\ptaa$. Earlier in Sect.~\ref{sec:simulations} we noticed that the contamination
estimates are biased towards high values with low SN observations. 
Not only does the higher contamination posterior limit increase as the SN
level decreases, but also the contamination mode moves from zero towards higher values.

\begin{figure}
	\centering
	\includegraphics[width=\columnwidth]{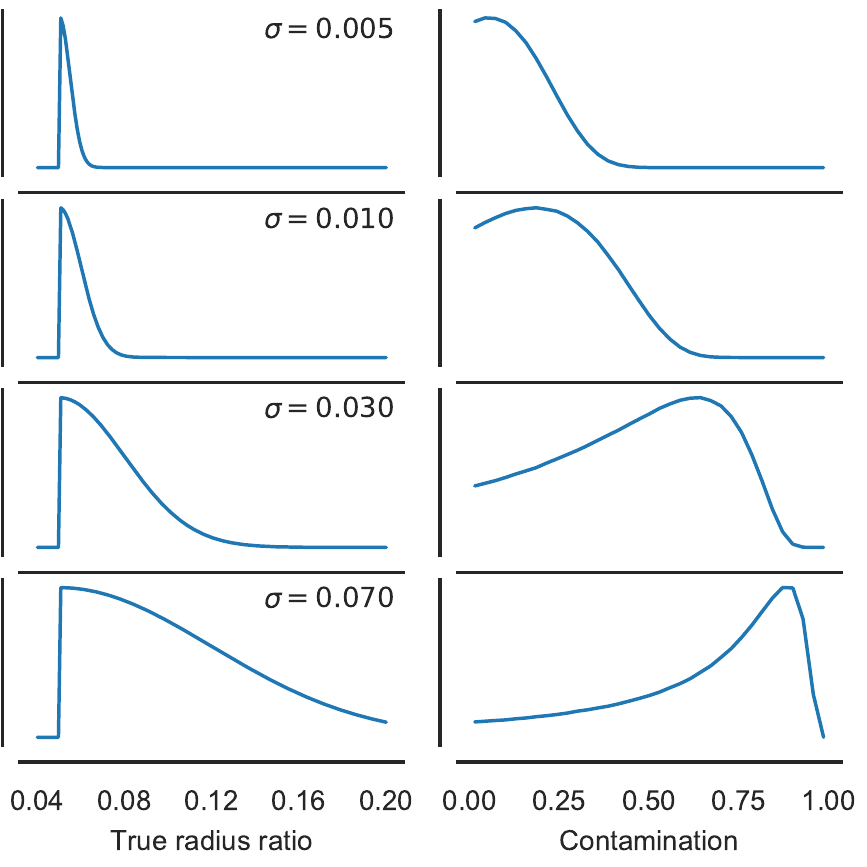}
	\caption{The effect of widening true radius ratio posterior distribution on the contamination posterior distribution. We assume a constant apparent radius ratio of 0.04, and model the true radius ratio posterior as a half-normal distribution with the mode at 0.04, and four values for the standard deviation. The panels on the right show the contamination distribution corresponding to the true radius ratio distribution on the left panel.}
	\label{fig:ktrue_to_contamination_sigma}
\end{figure}

This is due to the nonlinear relation between the contamination and the true radius ratio.
When $\ptra \sim \para$, a small change in \ptra leads to a large change in contamination.
Keeping the \ptra posterior mode constant, but increasing its width, moves the 
contamination posterior mode towards unity, as illustrated in Fig.~\ref{fig:ktrue_to_contamination_sigma}. Moving the \ptra posterior mode away from \para but keeping its width constant leads to 
the contamination posterior mode moving towards higher values with the width of the distribution
decreasing.

Thus, when interpreting the contamination posterior distribution, a high mode but long tail
means that the observations cannot constrain the true radius ratio well, while a high sharply
peaked mode means that the observations support strong contamination.

\begin{figure}
	\centering
	\includegraphics[width=\columnwidth]{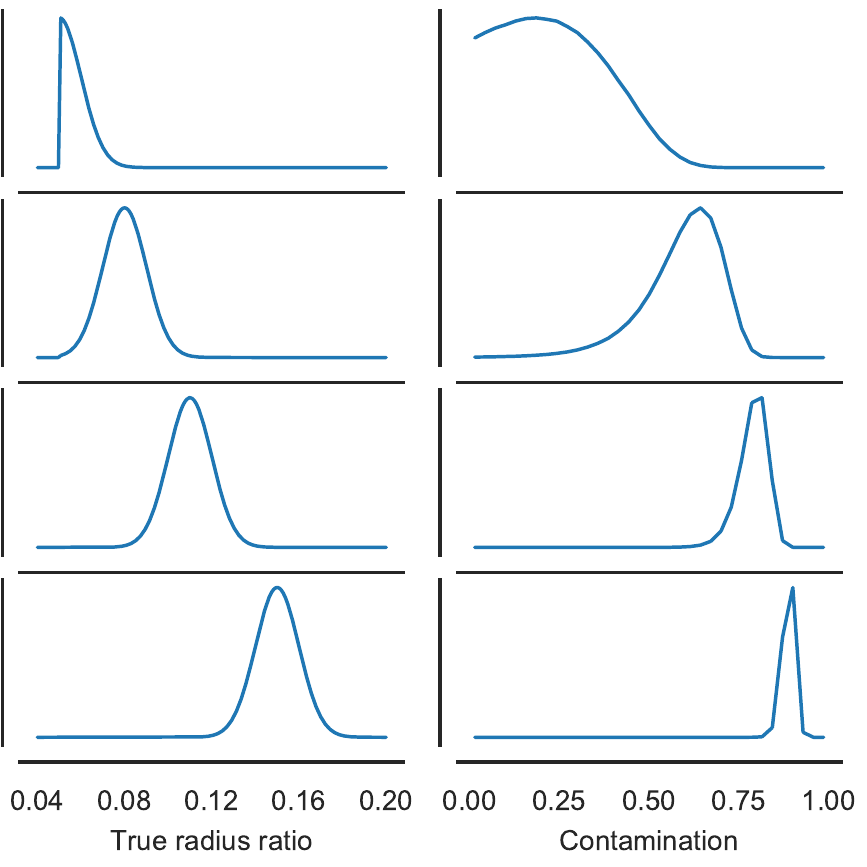}
	\caption{The effect of changing mode of \ptra posterior on the contamination posterior distribution. Apparent radius ratio is again assumed constant, $\para = 0.04$, and the \ptra posterior is modelled as concatenated normal distribution with a lower limit of 0.04. The panels on the right show the contamination distribution corresponding to the true radius ratio distribution on the left panel.}
	\label{fig:ktrue_to_contamination_centre}
\end{figure}

\end{document}